\documentclass{aa}
\usepackage[utf8]{inputenc}
\usepackage{graphicx}
\usepackage{epstopdf}
\usepackage{txfonts}
\usepackage{verbatim}
\usepackage{siunitx}
\usepackage{hyperref}
\hypersetup{colorlinks, allcolors=blue}
\usepackage{subcaption}
\usepackage{comment}
\usepackage[labelfont=bf]{caption}
\usepackage{tablefootnote}
\usepackage[flushleft]{threeparttable}
\usepackage{caption}
\usepackage{amsmath}
\usepackage{relsize}
\usepackage{multirow}
\usepackage{mathtools}
\usepackage{orcidlink}

\usepackage[switch]{lineno}

\begin{document}

\title{The ESO SupJup Survey}
\subtitle{X. A carbon isotope contrast in the young ROXs 12 system} 
\titlerunning{SupJup X: ROXs 12A and ROXs 12B}

\author{
N. Grasser\inst{\ref{instLeiden}}\orcidlink{0009-0009-6634-1741} \and 
I. A. G. Snellen\inst{\ref{instLeiden}}\orcidlink{0000-0003-1624-3667} \and 
S. de Regt\inst{\ref{instLeiden}}\orcidlink{0000-0003-4760-6168} \and 
D. Gonz\'alez Picos\inst{\ref{instLeiden}}\orcidlink{0000-0001-9282-9462} \and
Y. Zhang\inst{\ref{instCalTech}}\orcidlink{0000-0003-0097-4414} \and
T. Stolker\inst{\ref{instLeiden}}\orcidlink{0000-0002-5823-3072} \and
S. Gandhi\inst{\ref{instWarwick},\ref{instCEH}}\orcidlink{0000-0001-9552-3709} \and
R. Landman\inst{\ref{instLeiden}}\orcidlink{0000-0002-7261-8083} \and
P. Mollière\inst{\ref{instMPI}}\orcidlink{0000-0003-4096-7067}
\and
N. F. Allard\inst{\ref{instParis1},\ref{instParis2}}\orcidlink{0000-0001-6220-6221}
}

\institute{
Leiden Observatory, Leiden University, Postbus 9513, 2300 RA, Leiden, The Netherlands \\
\email{grasser@strw.leidenuniv.nl} \label{instLeiden}
\and 
Department of Astronomy, California Institute of Technology, Pasadena, CA 91125, USA \label{instCalTech} \and
Department of Physics, University of Warwick, Coventry CV4 7AL, UK \label{instWarwick} \and
Centre for Exoplanets and Habitability, University of Warwick, Gibbet Hill Road, Coventry CV4 7AL, UK \label{instCEH} \and
Max-Planck-Institut für Astronomie, Königstuhl 17, 69117 Heidelberg, Germany \label{instMPI} \and
LIRA, Observatoire de Paris, Universit\'e PSL, Sorbonne Universit\'e, Sorbonne Paris Cit\'e, CNRS, 61, Avenue de l'Observatoire, F-75014 Paris, France \label{instParis1} \and
Institut d'Astrophysique de Paris,  UMR7095, CNRS, Universit\'e Paris VI, 98bis Boulevard Arago, F-75014 PARIS, France \label{instParis2}
}
\date{}

\abstract{Emerging research suggests that elemental and isotopic ratios of exoplanet and brown dwarf atmospheres may serve as potential tracers of their formation pathways. The ESO SupJup Survey aims to shed light on this hypothesis, with a focus on the $^{12}$CO/$^{13}$CO ratio, by investigating the atmospheric composition of substellar companions and isolated brown dwarfs.
} 
{In this work, we aim to characterize the atmospheres and determine the ratios of $^{12}$CO/$^{13}$CO of the Rho Ophiuchus X-ray source (ROXs) 12 system ($\sim$\,6\,Myrs), consisting of an M0 host with an L0 companion, as part of the ESO SupJup survey. This system provides a great opportunity to directly compare the atmospheric compositions of the host star and its companion.} 
{Using high-resolution CRIRES+ K band spectra of these objects, we perform atmospheric retrieval analyses to derive their atmospheric properties, including the $^{12}$CO/$^{13}$CO ratio. Our retrieval framework is built on the radiative transfer code \texttt{petitRADTRANS}, with which we generate model spectra based on equilibrium chemistry tables computed with \texttt{FastChem}, coupled with the nested sampling algorithm \texttt{PyMultiNest}.}
{We report the presence of H$_2$O, $^{12}$CO, $^{13}$CO, and HF in both the star and companion, with a tentative detection of H$_2^{18}$O in ROXs~12B. The $^{12}$CO/$^{13}$CO ratios of the two objects show a measurable, though not strongly significant, difference, namely $77\,\substack{+10 \\ -7}$ and $55\,\substack{+10 \\ -7}$ for ROXs~12A and B. Both are consistent with the local present interstellar medium. We measure a C/O ratio of 0.54\,$\pm$\,0.01 and obtain a lower limit of H$_2^{16}$O/H$_2^{18}$O\,$\gtrapprox$\,300 for ROXs~12B, while the C/O ratio of the star is not reliably constrained due to the absence of atomic oxygen lines in the K band. The companion also appears to exhibit a more isothermal temperature structure than expected from models. Furthermore, we retrieve moderate veiling in the host star of $r_k$\,=\,$0.17\,\substack{+0.02 \\ -0.03}$.
} 
{
Systems such as ROXs 12, in which both star and planet can be chemically and isotopically characterized, are crucial for constraining potential formation mechanisms of massive, wide-orbit super-Jupiters. The differing $^{12}$CO/$^{13}$CO ratios in the ROXs~12 system highlight the need for a broader sample to assess the frequency of isotopic variations and whether they may be linked to formation history.
}


\maketitle

\section{Introduction}

Directly imaged substellar companions, which range from planets to brown dwarfs at distances of up to hundreds of AU, provide very favorable conditions for atmospheric studies. Although the sample size of these objects is currently rather small, the number of discoveries has been steadily increasing in the last decade (e.g., \citealt{Deacon2016, Kuzuhara2022}). Of these widely separated companions, planetary-mass objects have been argued to form in the disk via core accretion closer to the star and eventually become scattered to larger separations \citep{Boss2006}. However, the formation mechanisms of more massive substellar companions, such as giant exoplanets and companion brown dwarfs, are less clear \citep{Bergin2024}. They may also form through cloud collapse and fragmentation \citep{Boss2001, Bate2009}, or instabilities in massive protoplanetary disks \citep{Boss1997, Kratter2016}. In this regard, investigating atmospheric tracers is crucial for understanding and possibly distinguishing between these different formation processes. 

The composition, climate, and chemical and physical processes of exoplanets and brown dwarfs are encoded within their atmospheric spectra. Since the chemical composition of solids and gases is expected to depend on their birth environment, the chemical constituents of these objects could reveal their formation and evolutionary pathways \citep{Turrini2021, Pacetti2022}. Formation tracers include elemental ratios such as C/O (e.g., \citealt{Oberg2011}) and isotope ratios such as D/H and $^{12}$C/$^{13}$C (e.g., \citealt{Morley2019, MolliereSnellen2019}). For example, objects that formed via gravitational collapse are thought to retain the $^{12}$C/$^{13}$C ratio of their parent cloud \citep{Oberg2021}, which is expected for isolated brown dwarfs and stars \citep{Bate2002}. Conversely, objects that formed by core accretion are believed to inherit the $^{12}$C/$^{13}$C ratio of local disk solids, which can vary throughout the disk due to isotope fractionation processes \citep{Visser2009, Yoshida2022}.

To disentangle the formation pathways of super-Jupiters, substellar companions, and isolated brown dwarfs, the atmospheric properties of a large sample of these objects are analyzed as part of the ESO SupJup survey (Program ID: 1110.C-4264, PI: Snellen, \citealt{deRegt2024}). High-resolution K band and some J band spectra of 19 isolated objects, 19 lower-mass companions, and 11 hosts were obtained with the upgraded CRyogenic high-resolution InfraRed Echelle Spectrograph (CRIRES+) on the Very Large Telescope (VLT) at the Paranal Observatory in Chile (e.g., \citealt{Kaeufl2004, Follert2014, Dorn2014}). This work is part of the ongoing analysis of the data obtained throughout the survey, with the aim of characterizing the Rho Ophiuchus X-ray source (ROXs) 12 system, which consists of a stellar host and a substellar companion.

The star ROXs~12 was first discovered by \cite{Montmerle1983} during a study of the $\rho$ Ophiuchus star-forming region. \cite{Bouvier1992} determined that it is a very young M0 star, with its position in the Hertzsprung-Russell diagram indicating a subsolar mass and an age of a few Myr. While conducting a stellar multiplicity survey, \cite{Ratzka2005} identified a widely separated candidate companion for ROXs~12. This companion, denoted ROXs~12B, was finally confirmed years later by \cite{Kraus2014}, located at a separation of 1.7" from its host star. ROXs~12A appears to host a passive disk with no signs of accretion, and ROXs~12B likewise shows no evidence for active accretion, as indicated by the absence of Pa$\beta$ emission \citep{Bowler2017}. This is consistent with the non-detection of submillimeter continuum emission at 0.88\,mm at the location of the companion \citep{Wu2020}, which rules out the presence of a massive dust disk. Furthermore, the spectrum of ROXs~12B exhibits indications of low surface gravity \citep{Bowler2017}, such as shallower KI doublets and reduced FeH absorption \citep{Allers2013}. Using high-resolution spectroscopy from the Keck Planet Imager and Characterizer, \cite{Xuan2024} find values of C/O\,=\,0.54\,$\pm$\,0.05, a $0.5\substack{+0.4 \\ -0.2} \,\,\times$ solar metallicity ([C/H]\,=\,$-0.30$\,$\substack{+0.26 \\ -0.22}$, assuming a clear atmosphere), and $^{12}$CO/$^{13}$CO\,$\sim$\,100 for ROXs~12B, though the detection significance for the latter is below 3\,$\sigma$. For ROXs~12A, \cite{Swastik2021} report a metallicity of [Fe/H]$\,=\,0.14\,\pm\,0.01.$

\begin{table}[t!]
    \centering
    \caption{Properties of the ROXs 12 system from the literature.}
    \resizebox{0.97\linewidth}{!}{
    \begin{tabular}{lcc}
    \hline
    \hline
     Parameter & ROXs 12A & ROXs 12B  \\
     \hline
     Spectral type &  M0$^{(\text{a})}$ & L0$^{(\text{a})}$ \\
     Mass & $0.65\,\substack{+0.05 \\ -0.09}\,M_{\odot}^{(\text{a})}$ & $17.5\pm1.5\,M_\text{Jup}^{(\text{a})}$ \\
     Semi-major axis [AU] & -- & $210\pm20^{(\text{b})}$ \\
     Separation ["] & -- & $1.7^{(\text{b})}$ \\
     Distance [pc] & \multicolumn{2}{c}{$138.6\pm0.3^{(\text{c})}$} \\
     RA, Dec (ICRS, J2000) & \multicolumn{2}{c}{16:26:28.0397, $-$25:26:47.7175$^{(\text{c})}$} \\
     Radial velocity [km/s] &  
     $-5.8\pm1.9 ^{(\text{d})}$ & -- \\
     Rotational velocity [km/s] & $7.21 \pm 0.07^{(\text{d})}$ & 
     3.6$\substack{+1.2 \\ -1.6}^{(\text{e})}$ \\
     Age [Myr] & \multicolumn{2}{c}{6\,$\substack{+4 \\ -2}^{(\text{a})}$} \\
     $K_{\text{S}}$ [mag] & $9.10\pm0.03^{(\text{a})}$ & $14.14\pm 0.03^{(\text{a})}$ \\
     \([\text{Fe}/\text{H}]\) & 0.14\,$\pm\, 0.01^{(\text{f})}$ & $-0.30\,\substack{+0.26 \\ -0.22}^{(\text{e})}$\\
      Effective temperature [K] &
      $3850\,\substack{+100\\ -70}^{(\text{b})}$
      &  $2500\pm140^{(\text{e})}$ \\
     \hline
    \end{tabular}
    }
    \tablefoot{References: $^{(\text{a})}$\cite{Bowler2017}, $^{(\text{b})}$\cite{Kraus2014},$^{(\text{c})}$\cite{Gaia2020}, $^{(\text{d})}$\cite{Bowler2023}, 
    $^{(\text{e})}$\cite{Xuan2024}, $^{(\text{f})}$\cite{Swastik2021}}
    \label{tab:properties}
\end{table}

In this work, the atmospheres of the stellar host and brown dwarf companion in the ROXs~12 system are analyzed. We apply an atmospheric retrieval framework to their observed high-resolution spectra to infer their atmospheric properties. The observational setup and spectral extraction are outlined in Section~\ref{sec:observations}. Section~\ref{sec:methods} describes our applied retrieval framework, which encompasses atmospheric modeling and application of a nested sampling tool. In Section~\ref{sec:results} and Section~\ref{sec:discuss}, we report and discuss our findings, concluding our work in Section~\ref{sec:conclusion}.

\section{Observations}\label{sec:observations}

\begin{figure}
    \centering
    \includegraphics[width=0.95\linewidth]{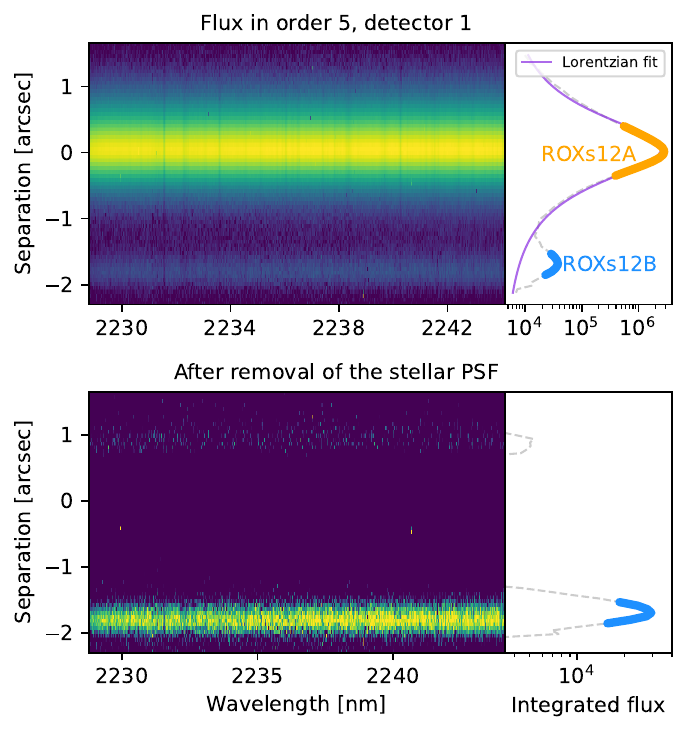}
    \caption{Calibrated observational data of order 5, detector 1, nodding position A. The x-axis on the left panels shows the wavelength in nanometers, while the separation from the star in arcseconds is shown on the y-axis. The bright signal in the center is the star ROXs~12A. Its companion ROXs~12B is seen as the faint source at -1.7~arcsec. The integrated flux as a function of the separation is shown in the right panels, where the extraction range of the star and companion are highlighted. The Lorentzian fit to the stellar PSF is shown in purple, which is used to remove the starlight, as shown in the bottom panel.
    }
    \label{fig:observation}
\end{figure}

\begin{table}[t!]
    \centering
    \caption{Setup and conditions during the observation of the ROXs 12 system.}
    \resizebox{0.7\linewidth}{!}{
    \begin{tabular}{lc}
    \hline
    \hline
    Observation date & 2023-03-05 \\
    Wavelength setting & K2166 \\
     Slit width ["] & 0.4 \\
     Exposure time & 3\,h\,\,40\,min \\
     Seeing ["] & 0.8\,$\pm$\,0.07 \\
     Airmass & 1.7--1.0\\
     Integrated water vapor [mm] & 7.99\,$\pm$\,0.21 \\
     Resolution & 60 000 \\
     \hline
    \end{tabular}
    }
    \label{tab:observations}
\end{table}

We obtained high-resolution K band spectra of the ROXs~12 system on March 5, 2023, with the CRIRES+ instrument on the VLT. A slit width of 0.4" was used, with both ROXs~12A and ROXs~12B being observed simultaneously. The observations were carried out in the K2166 wavelength setting, covering the wavelength range of 1.92 to 2.48\,$\mu$m. The total exposure time of 3\,h 40\,min was split into 22 exposures of 600\,s each, utilizing an ABBA nodding sequence with a 4.5" nod throw along the slit. No adaptive optics were used due to high humidity, with the integrated water vapor being 7.99\,$\pm$\,0.21\,mm on average during our observations. The observing conditions were relatively stable, with an average seeing of 0.8\,$\pm$\,0.07", although the air mass changed considerably from 1.7 at the beginning to 1.0 at the end of the observations. Table~\ref{tab:observations} lists the main parameters of our observing setup and conditions. To help remove the atmospheric features of Earth  from our data, the telluric standard star i~Sco was observed immediately before ROXs~12, at an air mass of roughly 1.9 to 1.8.

The observations were reduced using \texttt{excalibuhr}\footnote{\url{https://github.com/yapenzhang/excalibuhr}} \citep{Zhang2024}, a Python package for reducing CRIRES+ spectroscopy implementing methods from \cite{HolmbergMadu2022} and \texttt{pycrires} \citep{StolkerLandman2023}. The pipeline includes flat-fielding the exposures to account for pixel-to-pixel variations, tracing the spectral orders as well as their curvature, extracting the blaze function, and removing the sky background through nodding pair subtraction. We used the wavelength solution of the standard star as our initial wavelength grid for the data. To obtain the spectra, we applied the optimal extraction methodology from \cite{Horne1986} using a Gaussian spatial profile. The extraction ranges were chosen to obtain a favorable balance between the S/N and contrast. We find that using a narrower extraction range for the companion is essential to avoid contamination from the star. Modest changes in the extraction apertures do not significantly affect the spectral shape or line depths, as the companion signal is dominated by the central pixels of the point spread function (PSF), while the wings contribute little additional information but increase noise and potential contamination. We set a half-aperture of 5 pixels for ROXs~12A and 3 pixels for ROXs~12B, corresponding to 0.295" and 0.177" on the sky.

The top left panel of Figure~\ref{fig:observation} shows a section of the calibrated observations of the ROXs~12 system. The bright central signal is from the host star, while the fainter signal below is from the companion. Telluric absorption features can be seen as vertical lines. To remove the starlight for the extraction of the companion, we found that a Lorentzian fit was best able to reproduce the stellar PSF at the position of the companion. We fit a Lorentzian profile to the host star's integrated flux in each order-detector pair, which we use to scale and subtract the stellar spectrum accordingly. The integrated flux in the top right panel of Figure~\ref{fig:observation} highlights the extraction range (central index with the half-aperture extent) of the star and companion in their corresponding colors, along with the Lorentzian fit to the stellar PSF. In the bottom panel, we show the flux after removal of the stellar PSF.

The extracted spectra have a median per-pixel S/N of approximately 13 for the companion star and 112 for the host star at 2345\,nm. We refined the wavelength solution using the wavelength optimization routine from \texttt{excalibuhr}, which adjusts the wavelength solution through a quadratic polynomial correction, obtained by maximizing the cross-correlation functions between the data and a telluric template. Only minor corrections up to $\pm$\,0.003\,nm, corresponding to 0.4 pixels, were required, which is well below one resolution element. Before combining the spectra extracted from the individual nodding positions, we divided each order-detector pair by the corresponding blaze function extracted with \texttt{excalibuhr}. Furthermore, we identified $\lesssim$\,100 problematic pixels by taking the difference between the two nodding positions, which reveals outliers that occur only in one nodding position, such as cosmic rays. As the detector edges tend to be more problematic, we also masked the first and last 10 pixels of each detector. We then combined the spectra of the two nodding positions by interpolating the flux of nodding position B onto the wavelength grid of nodding position A and subsequently adding them.

To fit the telluric features of Earth’s atmospheric transmission, we applied the software \texttt{Molecfit} \citep{Smette2015} to the observations of the standard star, which generates a synthetic telluric model based on the atmospheric conditions. Dividing our data by this model yields the final telluric-corrected spectra. Due to the significant changes in air mass during our observations, tellurics are particularly challenging to accurately correct for. Before applying the model, we scaled its air mass to match the science data. After testing several scaling factors, we find that the correction performs best when adopting an air mass slightly higher than the mean of the observations (1.4 rather than 1.22). The high humidity further enhances the strength of the telluric absorption, making certain wavelength regions especially difficult to correct. Therefore, we masked pixels containing the deepest tellurics, where the flux is below 60\% of the normalized continuum, as well as the region of 30 pixels around them. As a result, the entire bluest order is masked out. The instrumental throughput is recovered by dividing the blackbody spectrum of the standard star i~Sco, namely $T_{\text{eff}}$\,=\,15~000\,K \citep{Royer2024}, to remove the continuum of the standard star from the telluric model.

The spectral data consists of seven spectral orders, each of which covers three detectors with 2048 pixels each. As the line profiles of the standard star were fit with a Gaussian, the resolving power $\mathcal{R}$ can be estimated through the full width at half maximum (FWHM). According to the telluric line fits, the spectral data exhibit a spectral resolution of $\mathcal{R}$\,=\,$60~000$, which surpasses the nominal resolution of CRIRES+ in the wide slit mode at $\mathcal{R}$\,=\,$50~000$, due to good seeing conditions.

\section{Retrieval framework}\label{sec:methods}

\begin{table*}[]
    \centering
    \caption{Free parameters, prior ranges, and retrieval results for ROXs~12A and ROXs~12B.}
    \resizebox{0.95\textwidth}{!}{
    \begin{tabular}{llcc|rr}
    \hline
    \hline
    Parameter & Description & \multicolumn{2}{c|}{Prior range} & \multicolumn{2}{c}{Retrieval results} \\
    & & ROXs~12A & ROXs~12B & ROXs~12A & ROXs~12B \\
    \hline
$v_{\text{rad}}$ [km/s] & Radial velocity & \multicolumn{2}{c|}{$\mathcal{U}$(-20,20)} & -5.46$^{+0.03}_{-0.03}$  & -5.28$^{+0.02}_{-0.02}$ \\
$v\,\text{sin}\,i$ [km/s] & Projected rotational velocity & \multicolumn{2}{c|}{$\mathcal{U}$(0,40)} & 8.37$^{+0.05}_{-0.05}$  & 2.14$^{+0.21}_{-0.16}$ \\
log $g$ [cm/s$^2$] & Surface gravity & $\mathcal{G}$(4.1,0.2) & $\mathcal{G}$(4.0,0.2) &  4.45$^{+0.04}_{-0.04}$  & 4.03$^{+0.04}_{-0.04}$ \\
$\epsilon_\mathrm{limb}$ & Limb-darkening coefficient & \multicolumn{2}{c|}{$\mathcal{U}$(0.2,1)} & 0.63$^{+0.05}_{-0.05}$  & 0.60$^{+0.21}_{-0.25}$ \\
\hline
C/O & Carbon-to-oxygen ratio & \multicolumn{2}{c|}{$\mathcal{U}$(0.1,1)} &  0.87$^{+0.01}_{-0.01}$  & 0.54$^{+0.01}_{-0.01}$ \\
$[$Fe/H$]$ & Metallicity & \multicolumn{2}{c|}{$\mathcal{U}$(-1,1)} &  -0.06$^{+0.02}_{-0.03}$  & -0.53$^{+0.03}_{-0.03}$ \\
log $^{12}$CO/$^{13}$CO & log$_{10}$ ratio of $^{12}$CO to $^{13}$CO & \multicolumn{2}{c|}{$\mathcal{U}$(1,6)} & 1.89$^{+0.06}_{-0.05}$  & 1.74$^{+0.08}_{-0.07}$ \\
log H$_2^{16}$O/H$_2^{18}$O & log$_{10}$ ratio of H$_2^{16}$O to H$_2^{18}$O & \multicolumn{2}{c|}{$\mathcal{U}$(1,6)} & 4.34$^{+0.90}_{-1.08}$  & 2.53$^{+1.17}_{-0.33}$ \\
\hline
$T_0$ [K] & Temperature at $P_0=10^2\,$bar& $\mathcal{U}$(1e3,3e4) & $\mathcal{U}$(1e3,1e4) & 27797.57$^{+1064.09}_{-1282.16}$  & 3445.75$^{+160.55}_{-140.52}$ \\
$\nabla T_0$ & Gradient at $P_0=10^2\,$bar& \multicolumn{2}{c|}{$\mathcal{U}$(0,0.4)} & 0.30$^{+0.03}_{-0.03}$  & 0.03$^{+0.03}_{-0.02}$ \\
$\nabla T_1$ & Gradient at $P_1=10^0\,$bar& \multicolumn{2}{c|}{$\mathcal{U}$(0,0.4)} & 0.33$^{+0.01}_{-0.01}$  & 0.11$^{+0.01}_{-0.01}$ \\
$\nabla T_2$ & Gradient at $P_2=10^{-2}\,$bar& \multicolumn{2}{c|}{$\mathcal{U}$(0,0.4)} & 0.00$^{+0.01}_{-0.01}$  & 0.06$^{+0.01}_{-0.01}$ \\
$\nabla T_3$ & Gradient at $P_3=10^{-4}\,$bar& \multicolumn{2}{c|}{$\mathcal{U}$(0,0.4)} & 0.36$^{+0.02}_{-0.02}$  & 0.02$^{+0.01}_{-0.01}$ \\
$\nabla T_4$ & Gradient at $P_4=10^{-6}\,$bar& \multicolumn{2}{c|}{$\mathcal{U}$(0,0.4)} & 0.15$^{+0.10}_{-0.08}$  & 0.26$^{+0.07}_{-0.09}$ \\
\hline
$r_k(\lambda_\text{mid})$ & Veiling at 2166\,nm & \multicolumn{2}{c|}{$\mathcal{U}$(0,2)} & 0.17$^{+0.02}_{-0.03}$  & 0.02$^{+0.01}_{-0.01}$ \\
$T_{\mathrm{disk}}$ [K] & Disk temperature & \multicolumn{2}{c|}{$\mathcal{U}$(300,1500)} & 915.77$^{+97.46}_{-81.78}$  & 548.81$^{+114.11}_{-96.87}$ \\
log $\kappa_{\mathrm{cl},0}$ [cm$^2$/g] & Opacity at cloud base & -- & $\mathcal{U}$(-10,3) &  --  & -5.76$^{+2.24}_{-1.93}$ \\
log $P_{\mathrm{cl},0}$ [bar] & Cloud base pressure & -- & $\mathcal{U}$(-6,3) & --  & -0.68$^{+1.71}_{-2.07}$ \\
$f_{\mathrm{sed}}$ & Cloud decay power & -- & $\mathcal{U}$(0,20) & --  & 11.89$^{+3.99}_{-4.92}$ \\
\hline
log $a$ & GP amplitude & \multicolumn{2}{c|}{$\mathcal{U}$(-1,1)} & 0.95$^{+0.01}_{-0.01}$  & 0.17$^{+0.01}_{-0.01}$ \\
log $l$ [nm] & GP length-scale & \multicolumn{2}{c|}{$\mathcal{U}$(-3,0)} & -1.52$^{+0.01}_{-0.01}$  & -2.01$^{+0.01}_{-0.01}$ \\
    \hline
    \end{tabular}
    }
    \tablefoot{The three columns on the left show the parameter notation, description, and prior ranges. $\mathcal{U}$ represents a uniform prior within the given range, whereas $\mathcal{G}$ stands for a Gaussian prior with the given mean and standard deviation. The retrieval results for ROXs~12A and ROXs~12B, including their 1-$\sigma$ credible interval, are shown in the two columns on the right.}
    \label{tab:free_params}
\end{table*}

For the analysis of the extracted spectra, we applied the atmospheric retrieval framework described in our previous study \citep{Grasser2025}. Using the radiative transfer code \texttt{petitRADTRANS} (\texttt{pRT}; Version 2.7; \citealt{Molliere2019}), we generated model spectra based on the object's properties, which we retrieved as free parameters. We utilized the nested sampling tool \texttt{PyMultiNest} \citep{Buchner2014}, a Python wrapper of the Bayesian inference algorithm \texttt{MultiNest} \citep{Feroz2009, Feroz2019}, to iteratively sample the parameter space. A few notable changes were made to the framework used in \cite{Grasser2025}. For one, we continuum-normalized the data as well as the generated model spectra. We also placed Gaussian priors on the surface gravity and include veiling effects.

We list the free parameters in Table~\ref{tab:free_params}, including their descriptions, prior ranges, and retrieved values for ROXs~12A and B. Following \texttt{MultiNest} recommendations, we used importance nested sampling (INS) mode with a sampling efficiency of 5\% \citep{Feroz2019}. The retrievals were run with 500 live points until an evidence tolerance of 0.1 was reached.

\subsection{Model spectra}

The model spectra for our retrievals are generated with \texttt{pRT} according to the atmospheric properties which we retrieve as free parameters. In the sections below, we describe our modeling setup, following the methodology in \cite{Grasser2025}.

\subsubsection{Chemistry}

Collision-induced absorption from H$_2$--H$_2$ and H$_2$--He, as well as Rayleigh scattering of H$_2$ and He \citep{Dalgarno1962, Chan1965, Borysow1988} are considered continuum opacities in our models. We use the HITEMP line lists for $^{12}$CO and $^{13}$CO \citep{Li2015}, the ExoMol line lists for H$_2$O (POKAZATEL; \citealt{Polyansky2018}), H$_2^{18}$O \citep{Polyansky2017}, and HF \citep{Li2013, Coxon2015, Somogyi2021}. These species are included in the retrievals of both objects. For the host star, we additionally include OH and CN \citep{Wang2020, Tennyson2020}, as well as the atomic species Na \citep{Allard2019}, H, Ca, Ti, Si, Sc, and Fe \citep{Kurucz2003}. The H- opacity is included as implemented in \texttt{pRT} following \cite{Gray2008}.

To minimize the number of free parameters, we assume thermo-chemical equilibrium, which is especially advantageous given the large number of species included in the stellar model. We retrieve the C/O ratio, metallicity, and base-10 logarithm of the isotope ratios $^{12}$CO/$^{13}$CO and H$_2^{16}$O/H$_2^{18}$O as free parameters. We pre-compute a table of equilibrium abundances using \texttt{FastChem Cond} \citep{Kitzmann2024} where
the most extensive reaction networks and condensation are employed. The table is constructed in four dimensions: temperature, pressure, metallicity, and C/O (see Table~\ref{tab:eqchem_dims}). During the retrieval, linear interpolations are performed to obtain the volume mixing ratios (VMRs)
for the model atmospheres.

\begin{table}[t!]
    \centering
    \caption{Dimensions and steps of the chemical equilibrium table calculated with \texttt{FastChem Cond} \citep{Kitzmann2024}.}
    \label{tab:eqchem_dims}
    \begin{tabular}{c|c}
        \hline\hline
        $T\ [\mathrm{K}]$   & $150, 200, 250, ..., 6000$ \\
        $P\ [\mathrm{bar}]$ & $10^{-5.0}, 10^{-4.9}, 10^{-4.8}, ..., 10^{+3.0}$ \\
        $\mathrm{[Fe/H]}$ & $-1.0, -0.9, -0.8, ..., +1.0$ \\
        $\mathrm{C/O}$    & $0.10, 0.20, 0.30, ..., 1.00$ \\
        \hline
    \end{tabular}
\end{table}

\subsubsection{Thermal profile and clouds}

We parameterize the pressure-temperature ($P$--$T$) profile of the atmosphere through temperature gradients, following \cite{Zhang2023}. The atmospheric layers are defined along a log$_{10}$-uniform grid between $10^2$--$10^{-6}$~bar, with 50 layers in total. The temperature at the bottom of the atmosphere ($T_0$, at a pressure of $P_0$\,=\,$10^2$\,bar), as well as the temperature gradients $\nabla T_i$ at the pressure log$_{10}$($P$)\,=\,[-6, -4, -2, 0, 2]\,bar, are retrieved as free parameters. The gradients at the remaining atmospheric layers are quadratically interpolated from the five retrieved gradients. We calculate the temperature $T$ at the pressure point $P$ using the temperature gradient $\nabla T_i$ at each atmospheric layer $i$, where $i$ increases with altitude

\begin{equation}
    T_i = T_{i-1} \cdot \left( \frac{P_i}{P_{i-1}}  \right)^{\nabla T_i}  \quad \text{where}  \quad \nabla T_i = \frac{d \text{ln} T_i}{d \text{ln} P_i}.
\end{equation}

With an effective temperature of approximately 2500\,K, ROXs~12B is expected to host clouds composed primarily of refractory Al- and Ti-bearing condensates, such as grossite (CaAl$_4$O$_7$), hibonite (CaAl$_{12}$O$_{19}$), corundum (Al$_2$O$_3$), and calcium titanate (Ca$_3$Ti$_2$O$_7$) \citep{Wakeford2017}. To ensure unbiased atmospheric retrieval, we adopt a simple gray cloud treatment similar to that of \citet{Molliere2020}. The cloud opacity $\kappa_\mathrm{cl}(P)$ is set to $\kappa_{\mathrm{cl},0}$ at the cloud base $P_{\mathrm{cl},0}$, decays above the base with a power-law defined by $f_\mathrm{sed}$, and is zero below the base, defined by

\begin{equation}
\kappa_\mathrm{cl}(P) =
\begin{cases}
\kappa_{\mathrm{cl},0} \left(\dfrac{P}{P_\mathrm{base}}\right)^{f_\mathrm{sed}} & P < P_\mathrm{base} \\
0 & P \geq P_\mathrm{base}
\end{cases}
\end{equation}

\subsubsection{Surface gravity}

\cite{Bowler2017} determine a radius of 1.14\,$\pm$\,0.07\,$R_\odot$ and mass of $0.65\,\substack{+0.05 \\ -0.09}\, M_\odot$ for ROXs~12A, implying a surface gravity of log\,$g$\,$\sim$\,4.1, which agrees with predictions by evolutionary tracks, considering the system's age of $\sim$\,6\,Myr \citep{Baraffe2015}. Similarly, the mass of ROXs~12B (17.5\,$\pm$\,1.5\,$M_\text{Jup}$), \citealt{Bowler2017}) would imply a log\,$g$\,$\sim$\,4.0 when interpolating from evolutionary models \cite{Baraffe2015, Phillips2020}. To ensure physical solutions and minimize degeneracies associated with surface gravity, we adopt a Gaussian prior on log\,$g$ centered on 4.1 and 4.0 for A and B, with a standard deviation of 0.2\,dex, similar to the prior derived by \cite{Xuan2024} for ROXs~12B. These priors are justified considering that K band spectra lack gravity sensitive features \citep{Zhang2021_BD}, and the limited wavelength coverage can bias the retrieved log\,$g$ and other global parameters \citep{Burningham2021}.

\subsubsection{Veiling}

Due to the youth of the system, a circum(sub)stellar disk may be present, which reduces the depth of its photospheric absorption lines. We include veiling in the retrievals for both objects to test whether it can be constrained. Following \cite{Sullivan2019, Zhang2024}, the observed veiled spectrum $F'$ is expressed as an additive continuum excess

\begin{equation}
    F'(\lambda) = \frac{F(\lambda) + r_k(\lambda)F_0}{1+r_k(\lambda)},
\end{equation}

where $F$ is the intrinsic stellar spectrum and $F_0$ is the median flux level of $F$. We parameterize the veiling coefficient $r_k$ through a physically motivated approach, using the Planck function $B(\lambda,T_\text{disk})$ to model the wavelength dependence \citep{Antoniucci2017, Alcala2021}

\begin{equation}
    r_k(\lambda) = r_k(\lambda_\text{mid}) \frac{B(\lambda,T_\text{disk})}{B(\lambda_\text{mid},T_\text{disk})}.
\end{equation}

The temperature $T_\text{disk}$ sets the wavelength dependence of the veiling through the blackbody flux $B({\lambda},T_\text{disk})$, whereas $r_k(\lambda_\text{mid})$ specifies the veiling at the median of our wavelength range, namely $\lambda_\text{mid}$\,=\,2166\,nm, as our midpoint for normalization.

\subsubsection{Broadening and normalization}\label{subsec:norm}

After generating the model spectrum, the \texttt{fastRotBroad} function from \texttt{PyAstronomy}\footnote{\url{https://github.com/sczesla/PyAstronomy}} \citep{Gray2008, Czesla2019} is applied to broaden it according to the projected rotational velocity $v\,\text{sin}\,i$ and the limb-darkening coefficient $\epsilon_\text{limb}$. The model spectra are down-convolved to the same resolution as the data.

We continuum-normalize the data and the model spectra, as we noticed a slope in the residuals between the data and a preliminary best-fit model of ROXs~12A. We suspect that this may be caused either by a residual instrumental slope not reflected in the blaze function extracted by \texttt{excalibuhr} or potential extinction effects. To ensure that we treat the data and model in the same way, we found that the most robust method was to use Fourier filtering as implemented in \texttt{np.fft}\footnote{\url{https://numpy.org/doc/stable/reference/routines.fft.html}} to reconstruct the continuum from the lowest frequencies (using frequencies below 1/150 of the array length, i.e., 13 pixels for each order-detector pair of 2048 pixels). We found that smaller cutoffs began to distort line wings, whereas larger cutoffs did not sufficiently remove residual slopes. The Fourier filtering procedure is applied identically to both the data and the model spectra. After masking the same pixels in the model as in the data, both spectra are divided by their respective continua prior to log-likelihood evaluation.

\subsection{Log-likelihood}

The likelihood of the models is determined following the methodology in \cite{Ruffio2019}. For each order-detector pair, the log-likelihood is calculated as

\begin{equation} \label{equ:lnL}
\begin{split}
\text{ln}\mathcal{L} = -\frac{1}{2} (&\text{ln}(|\vec{\Sigma}|) + \text{ln}(\vec{m}^T \vec{\Sigma}^{-1} \vec{m}) \\
& + (N_d - N_\phi + \alpha -1) \cdot \text{ln} \underbrace{(\vec{R}^T \vec{\Sigma}^{-1} \vec{R})}_{\chi_0^2}),
\end{split}
\end{equation}

where $\vec{R}$ refers to the residuals between the data and the model spectrum, and $\vec{\Sigma}$ represents the covariance matrix, consisting of the flux uncertainties and the potential correlation between pixels. $N_d$ is the number of valid data points. Following \cite{Ruffio2019}, we set the normalization term $\alpha= 2$ and the number of linear parameters $N_\phi =1$. The log-likelihood values of each order-detector are subsequently summed to obtain the total log-likelihood of the model. Following the methodology described in \cite{deRegt2024}, we also include an uncertainty-scaling parameter to account for potential over- or underestimation of the flux uncertainties.

To account for correlated noise, we use the methodology described in \cite{deRegt2024}, based on \cite{Kawahara2022}, in which the off-diagonal elements of the covariance matrix are determined through Gaussian processes (GP). For this, we introduce two additional free parameters to model a Gaussian with the amplitude $a$, which scales the uncertainty, and a length-scale $\ell$, which sets the contribution of off-diagonal elements. We refer to \cite{deRegt2024, Grasser2025} for more details.

\subsection{Test on injected spectrum}

To assess the robustness of our spectral extraction and subsequent retrieval, we generate a \texttt{pRT} test spectrum with selected properties similar to the retrieved values of ROXs~12B. Specifically, we set a solar metallicity, C/O\,=\,0.55, log\,$g$\,=\,4, log$_{10}$($^{12}$CO/$^{13}$CO)\,=\,1.7, and log$_{10}$(H$_2^{16}$O/H$_2^{18}$O)\,=\,2.5, and a Sonora Bobcat $P$--$T$ profile of $T_\text{eff}$\,=\,2200\,K. We also include modest veiling at $r_k(\lambda_\text{mid})$\,=\,0.2 with a disk temperature of $T_\text{disk}$\,=\,1000\,K. We then inject this spectrum on the other side of the stellar PSF at -1.7\,arcsec (see Fig.~\ref{fig:obs_inj}) and extract it using the same procedure as for the real companion spectrum. The retrieval is performed on this test spectrum with an identical setup to that used for ROXs~12B.

\section{Results}\label{sec:results}

\begin{figure}
    \centering
\includegraphics[width=0.95\linewidth]{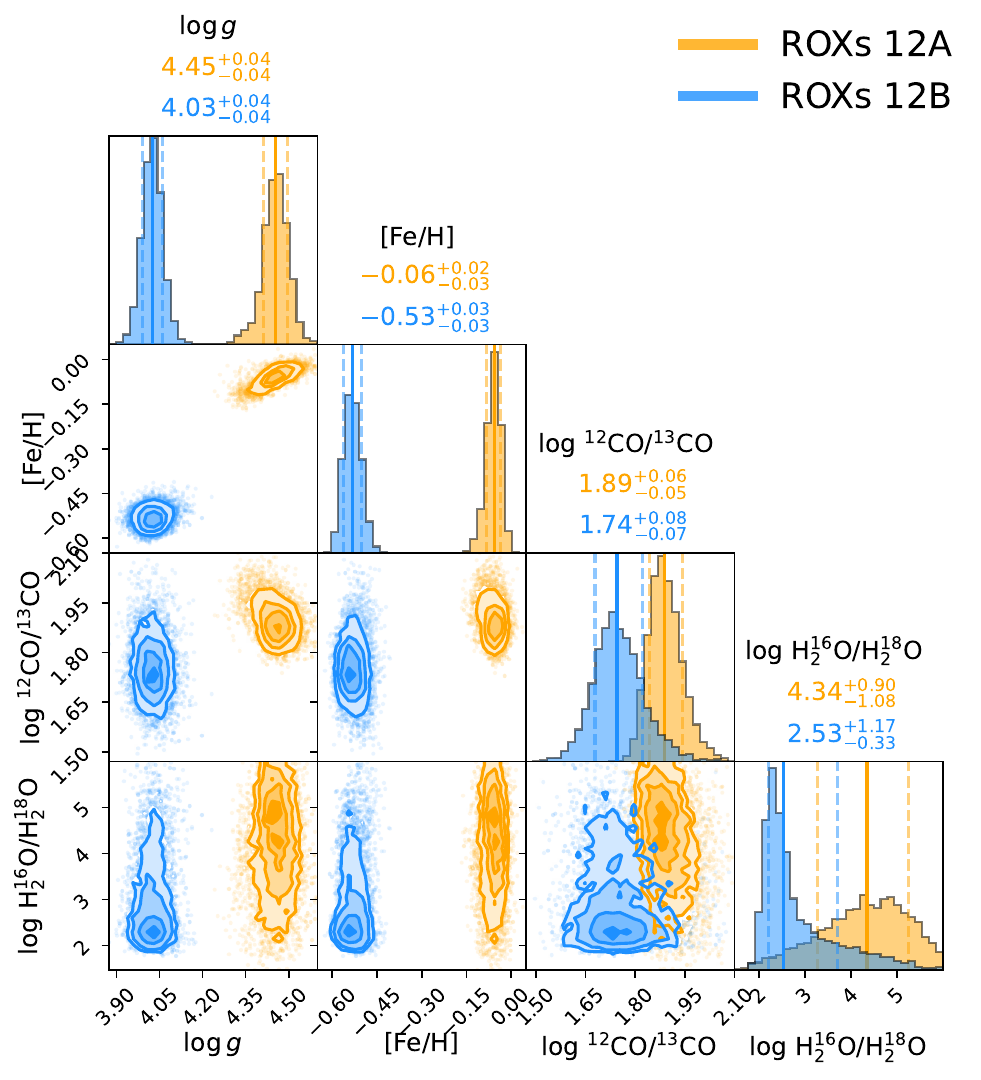}
    \caption{Posterior distributions of log\,$g$, [Fe/H], $^{12}$CO/$^{13}$CO, and H$_2^{16}$O/H$_2^{18}$O ratio for ROXs~12A (yellow) and B (blue).}
    \label{fig:cornerplot}
\end{figure}

\begin{figure*}
    \centering
\includegraphics[width=0.95\linewidth]{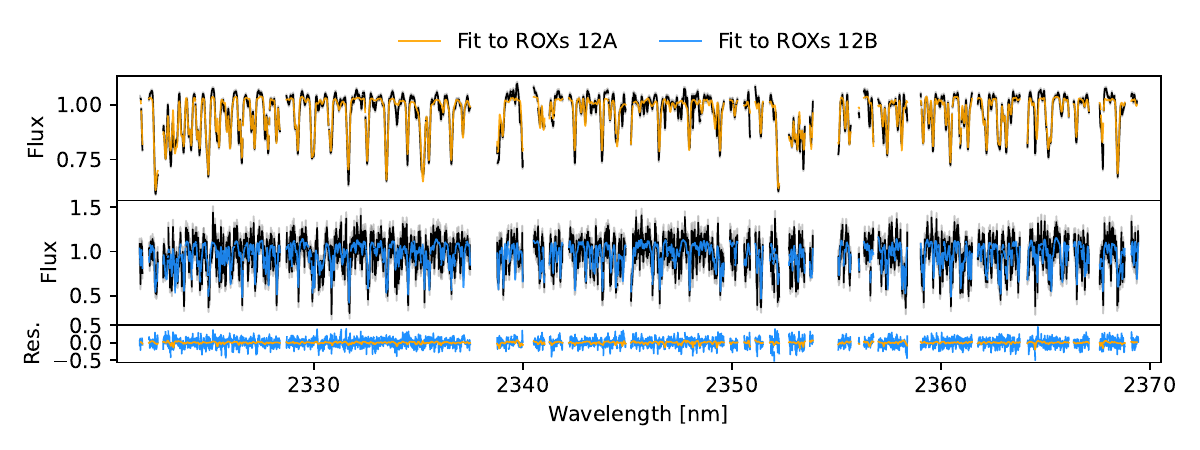}
    \caption{Best-fit models to ROXs~12A (top) and B (middle), with the residuals (bottom; data minus model), within a CO-dominated spectral order.}
    \label{fig:fit_CO}
\end{figure*}

\begin{figure}
    \centering
\includegraphics[width=0.95\linewidth]{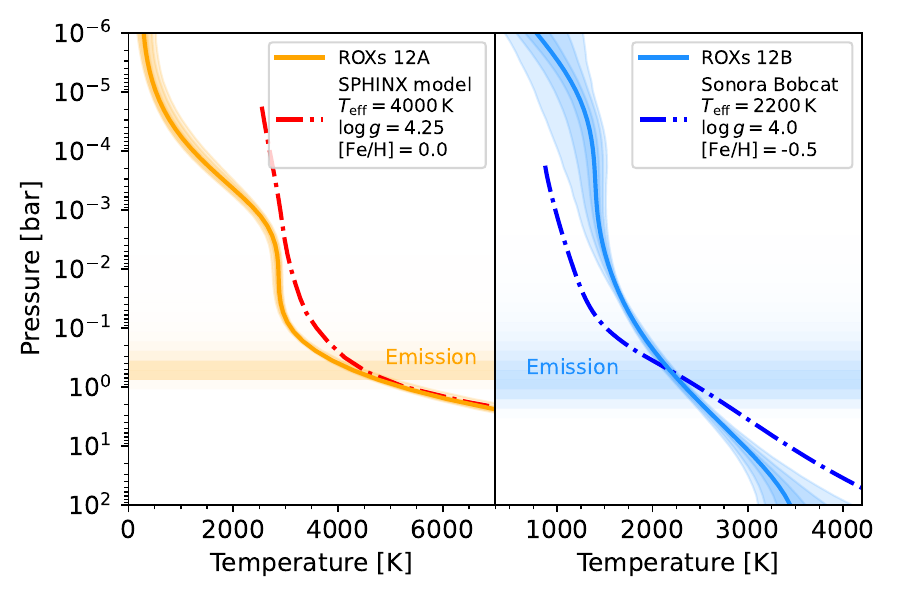}
    \caption{Retrieved $P$--$T$ profiles for ROXs~12A (left) and B (right). We compare ROXs~12A to a SPHINX model profile with $T_{\text{eff}}$\,=\,4000\,K, log\,$g$\,=\,4.25, [Fe/H]\,=\,0.0, and C/O\,=\,0.5 \citep{Iyer2023_SPHINX}, and ROXs~12B to a
    Sonora Bobcat model with $T_{\text{eff}}$\,=\,2200\,K, log\,$g$\,=\,4.0, [Fe/H]\,=\,-0.5, and solar C/O \citep{Marley2021_SonoraBobcat}.}
    \label{fig:PT}
\end{figure}

Table~\ref{tab:free_params} lists the parameter values retrieved  for ROXs~12A and B, with the posterior distributions of the key parameters shown in Fig.~\ref{fig:cornerplot}. We present a section of the data and the best-fit model spectra within a CO-dominated spectral order of ROXs~12A and B in Figure~\ref{fig:fit_CO}. The lower S/N of ROXs~12B can be clearly recognized in the larger residuals in the bottom panel. Overall, the fit reproduces the observed spectra well, with no noticeable residual slopes or systematic features. The full model spectra can be found in Appendix~\ref{app:bestfit}. We also show a summary of the results for the injected test spectrum in Fig.~\ref{fig:summary_inj}.

We note that some of the small parameter uncertainties may be overly optimistic, as narrow posteriors are a well-known limitation of \texttt{MultiNest} retrievals. In particular, when running \texttt{MultiNest} on constant efficiency mode, the sampler tends to focus on the high-likelihood core of the posterior and can underexplore the tails, leading to underestimated uncertainties rather than a fully representative posterior distribution. Higher-dimensional or noisy data may be especially susceptible, as well as parameters affected by degeneracies, weak constraints, or multimodal posterior structure (e.g., \citealt{Buchner2016, Lemos2023, Dittmann2024}).

\subsection{Thermal profiles}

We present the retrieved $P$--$T$ profiles for both objects of the ROXs~12 system in Fig.~\ref{fig:PT}. For ROXs~12A, the profile retrieved closely follows the SPHINX model corresponding to $T_{\text{eff}}$\,=\,4000\,K, log\,$g$\,=\,4.25, [Fe/H]\,=\,0.0, and C/O\,=\,0.5 \citep{Iyer2023_SPHINX} in the photospheric layers probed by our data. However, our $P$--$T$ profile deviates from the SPHINX model at lower atmospheric pressures, since our data are not sensitive to those pressure regions.

The profile retrieved for ROXs~12B is slightly cooler than $T_{\text{eff}}$\,=\,2500\,$\pm$\,140 found by \cite{Xuan2024}, both of which are markedly cooler than $T_{\text{eff}}$\,=\,3100\,$\substack{+400\\-500}$\,K reported by \cite{Bowler2017}. Our $P$--$T$ profile visually aligns more closely with Sonora Bobcat models \citep{Marley2021_SonoraBobcat} with $T_{\text{eff}}$\,$\approx$\,2200\,K. However, the effective temperature inferred from broadband spectroscopy does not necessarily correspond to the local photospheric temperature at the wavelengths probed, and the \cite{Bowler2017} $T_{\text{eff}}$ estimate carries substantial uncertainties. Furthermore, \cite{Bowler2017} themselves claim that their derived $T_{\text{eff}}$ is unusually high for an L0 dwarf and quote $T_{\text{eff}}$\,$\approx$\,2260\,$\pm$\,60\,K for ROXs~12B when instead using the $T_{\text{eff}}$-spectral type relation from \cite{Filippazzo2015}, which agrees with our retrieved photospheric temperature. 

Compared to the Sonora Bobcat $P$--$T$ profile \citep{Marley2021_SonoraBobcat}, we obtain shallower temperature gradients, with cooler temperatures at higher pressures and warmer temperatures at lower pressures than predicted. This behavior, also seen in other L-dwarf retrievals \citep{Burningham2017, Kitzmann2020, Lueber2022}, could be explained by clouds, although our retrieval is unable to constrain any cloud properties, or possibly chemical instability affecting the adiabatic index \citep{Tremblin2015, Tremblin2016}.

\subsection{Chemical composition}
 
As shown in Fig.~\ref{fig:cornerplot}, we obtain a roughly solar metallicity of ROXs~12A at $-0.06$\,$\substack{+0.02 \\ -0.03}$, while \cite{Swastik2021} suggest a slightly supersolar metallicity of 0.14\,$\pm$\,0.01. The metallicity we find for ROXs~12B at $-0.53$\,$\pm$\,0.03 is consistent with $-0.30\,\substack{+0.26 \\ -0.22}$ reported by \cite{Xuan2024}, who report similarly subsolar values for several other substellar companions. While such low metallicities are unexpected for young systems \citep{Santos2008} and may be driven by degeneracies between log\,$g$, temperature structure, and cloud properties, several other retrieval studies of substellar companions have also reported subsolar metallicities \citep{Line2021, Xuan2024, Inglis2024}. However, because these analyzes are subject to similar degeneracies, it remains unclear whether these low values reflect true atmospheric composition.

Interestingly, the retrieval of the injected test spectrum also yields a subsolar metallicity of $-0.29\,\substack{+0.04 \\ -0.03}$ despite the input being solar (Fig.~\ref{fig:summary_inj}). Since veiling was included in the model but not recovered by retrieval, this suggests that at low S/N the method struggles to disentangle the effects of metallicity, modest veiling, and clouds on line depths. The small uncertainties place the retrieved metallicity significantly away from the true value, which decreases our confidence in the metallicity retrieved for ROXs~12B. Accurate retrievals of the metallicity may only be possible for objects unaffected by veiling. Consequently, it cannot be ruled out that the subsolar metallicity inferred for ROXs~12B may reflect such parameter degeneracies rather than the true composition. However, recent JWST studies of disks around similar young companions (e.g., \citealt{Cugno2024}) indicate typical disk temperatures of $\approx$\,500K, implying that veiling contributions in the K band are expected to be negligible and primarily affect longer wavelengths. This consideration strengthens the interpretation that the low metallicity may be intrinsic rather than driven by unaccounted for veiling.

Our results indicate a C/O ratio for ROXs~12B of 0.54\,$\pm$\,0.01, close to solar \citep{Asplund2021}, and in agreement with \cite{Xuan2024}, who find C/O\,=\,0.54\,$\pm$\,0.05. In contrast, we obtain a higher C/O ratio for ROXs~12A around 0.87\,$\pm$\,0.01. However, we caution that this value may not represent the true C/O ratio of the host star. Within the star's photospheric temperatures, the dissociation of H$_2$O shifts oxygen into OH and, more significantly, atomic oxygen, which becomes the second-largest oxygen reservoir after $^{12}$CO according to our equilibrium chemistry tables, with its abundance increasing markedly toward lower C/O. Although atomic oxygen does not exhibit spectral features in the K band, OH does. However, OH responds only weakly to changes in total oxygen abundance at these temperatures, limiting the sensitivity of our retrieval to the full oxygen budget and making it difficult to obtain a reliable estimate of the true C/O ratio \citep{Asplund2001}.

We retrieve a $^{12}$CO/$^{13}$CO ratio of $77\,\substack{+10 \\ -7}$ and $55\,\substack{+10 \\ -7}$ for ROXs~12A and B. Both ratios are consistent with the present-day local interstellar medium (ISM) within the 1\,$\sigma$ scatter ($^{12}$C/$^{13}$C$_\text{ISM}$\,=\,68\,$\pm$\,15, \citealt{Milam2005}), with the companion’s ratio somewhat lower and the host star’s somewhat higher. \cite{Xuan2024} report that their $^{12}$CO/$^{13}$CO posterior peaks near $\sim$\,100 for ROXs~12B, but their detection does not exceed a 3\,$\sigma$ significance level. Other young companions \citep{Zhang2024, Xuan2024, Picos2025} and young isolated brown dwarfs \citep{Picos2024, deRegt2026} likewise exhibit $^{12}$CO/$^{13}$CO ratios consistent with the ISM.

For ROXs~12B, we find a ratio of H$_2^{16}$O/H$_2^{18}$O\,=\,$337\,\substack{+4682 \\ -181}$. The large upper uncertainty is due to the tail of the posterior distribution being in log$_{10}$-space, as seen in Fig.~\ref{fig:cornerplot}. The posterior distribution shows a clear peak of H$_2^{16}$O/H$_2^{18}$O at $\approx$\,300, suggesting that there may be a slight enrichment of H$_2^{18}$O compared to the Sun (525\,$\pm$\,21 \citealt{Lyons2018}) and the ISM (557\,$\pm$\,30 \citealt{Wilson1999}). However, due to the very large uncertainty towards larger values, the inclusion of H$_2^{18}$O is not favored at a statistically significant level, therefore this ratio should be interpreted as a lower limit. Other retrievals of isolated brown dwarfs show a wide range of H$_2^{16}$O/H$_2^{18}$O ratios, with some being above the ISM \citep{Grasser2025}, some below \citep{Picos2024}, and some undetectable \citep{Mulder2025}.

The retrieval of the injected test spectrum recovers the C/O (0.51\,$\pm$\,0.01 vs. 0.55), $^{12}$CO/$^{13}$CO (1.64\,$\substack{+0.04 \\ -0.03}$ vs. 1.7), and H$_2^{16}$O/H$_2^{18}$O (2.68\,$\substack{+0.16 \\ -0.11}$ vs. 2.5) ratios reasonably well, with only small offsets from the input values (see Fig.~\ref{fig:summary_inj}). This suggests that these ratios for ROXs~12B are relatively robust and less affected by degeneracies, such as those between metallicity, log\,$g$, veiling, clouds, or temperature structure.

\begin{figure}[t!] 
     \centering
    \includegraphics[width=0.95\linewidth]{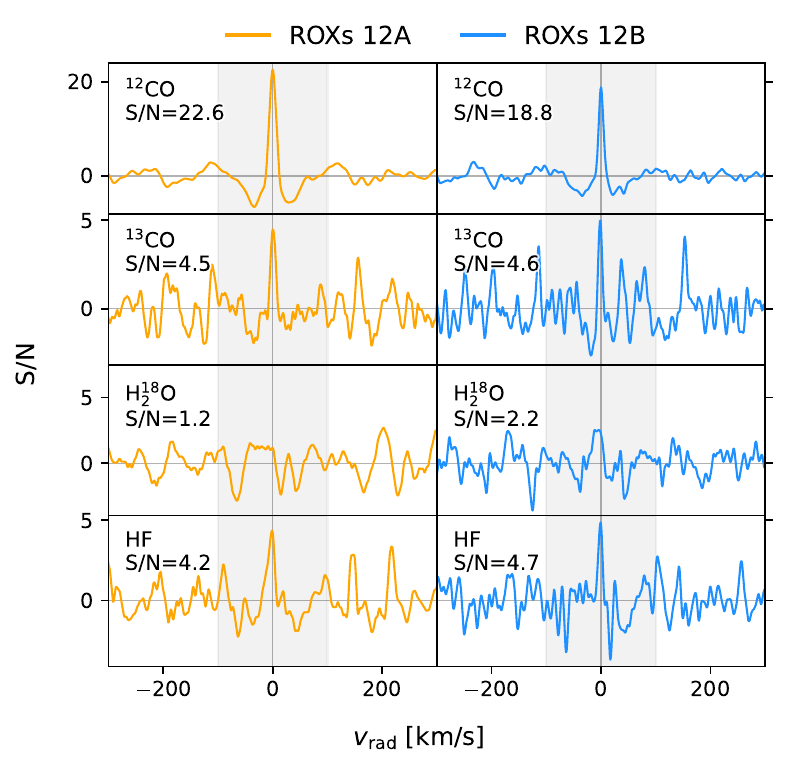}
     \caption{Cross-correlation functions of selected atmospheric species for ROXs~12A (left, yellow) and B (right, blue).}
     \label{fig:ccf}
\end{figure}

To further assess the robustness of the detections, we compute the cross-correlation functions (CCFs) for each atmospheric species. Following \cite{Zhang2021}, we compute the CCF between a template of the selected species, obtained by subtracting the fiducial model without that species from the full fiducial model, and the residuals, defined as the observed spectrum minus the fiducial model without the selected species. Fig.~\ref{fig:ccf} shows the CCFs for ROXs~12A (left) and B (right) of selected atmospheric species. We confirm the presence of $^{13}$CO and HF in both objects, while H$_2^{18}$O only produces a tentative signal in ROXs~12B, as expected from the retrieval posteriors in Fig.~\ref{fig:cornerplot}. This makes our result the first robust constraint of the $^{12}$CO/$^{13}$CO ratio in ROXs~12B. Interestingly, the detection confidence of $^{13}$CO is similar for both objects, despite the S/N of the stellar spectrum being significantly higher. This behavior may result from a combination of effects, including a lower intrinsic $^{13}$CO abundance in the star relative to the companion and the presence of modest veiling in the stellar spectrum, which reduces the observed line depths. Similar isotopic detection confidences of the host star and companion have also been found in other studies (e.g., \citep{Xuan2024b, Picos2025}).

Due to the low S/N of ROXs~12B, we were only able to include its most prominent absorbers in our retrieval, namely $^{12}$CO, $^{13}$CO, H$_2$O, H$_2^{18}$O, and HF. For ROXs~12A, on the other hand, we retrieve a wide range of species in addition to those retrieved for ROXs~12B, whose presence is confirmed by cross-correlation (see Fig.~\ref{fig:ccf_rest_A}). At these high temperatures, atomic hydrogen and H- are important constituents of chemical equilibrium. We also retrieved OH and CN, as well as several atomic species, namely Na, Ti, Fe, Ca, Si, and Sc.

Despite our efforts to identify as many species as possible in the spectrum of ROXs~12A, the origin of some features in the spectrum remains unexplained. Mismatches in spectral features and unidentified lines are common issues in M dwarf spectroscopy, often due to the effects of non-local thermodynamic
equilibrium in M dwarfs, magnetic fields, and potentially incomplete molecular opacities under these conditions (e.g., \citealt{Allard1995, Rajpurohit2018, Reggiani2019, Olander2021, Cristofari2022}). However, this does not affect our retrieval results for $^{12}$CO and $^{13}$CO, as the mismatched features occur mainly in the wavelength regions without CO absorption, while CO is the dominant contributor in the last two spectral orders.

\subsection{Veiling}

\begin{figure}[t!] 
     \centering
    \includegraphics[width=0.95\linewidth]{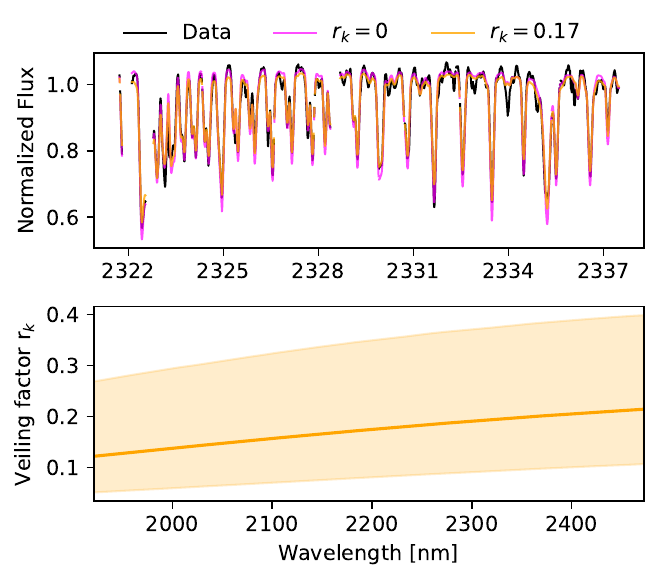}
     \caption{Top panel: Comparison of the best-fit model to ROXs~12A with (yellow) and without (magenta) veiling. Bottom panel: Veiling coefficient $r_k$ of ROXs~12A and its 1-$\sigma$ uncertainty range throughout the observed wavelength range.}
     \label{fig:veiling}
\end{figure}

We retrieve a veiling coefficient of $r_k$\,=\,$0.17\,\substack{+0.02 \\ -0.03}$ at 2166\,nm for ROXs~12A. The impact of the veiling is shown in the top panel of Fig.~\ref{fig:veiling}, where we compare it to a model without veiling. In the bottom panel, we show the wavelength-dependence of the veiling, which follows a $T$\,$\approx$\,900\,K blackbody. \cite{Bowler2017} also report disk emission for ROXs~12A, with an excess at 22\,$\mu$m with $T$\,$\approx$\,230\,K. However, this likely originates from cooler parts of the outer disk, whereas the veiling we retrieve seems to stem from the hot inner part of the disk. Hot inner-disk dust and gas are known to produce a near-infrared excess in the K band \citep{Muzerolle2003, Fischer2011}.

The retrieved veiling is modest but clearly non-zero. This is consistent with expectations for $\sim$6\,Myr systems, where veiling has typically declined relative to younger objects but often remains detectable. Stars with ages similar to ROXs~12A exhibit comparable near-infrared veiling,  with somewhat higher values found in more massive stars capable of sustaining larger disks
\citep{Sousa2023}. For example, TW Hydra, which has a similar age and spectral type, exhibits $r_k$\,$\approx$\,0.3--0.4 at 2.2\,$\mu$m \citep{Sokal2018}.

In contrast, we do not retrieve any veiling for the companion ROXs~12B, with a veiling coefficient of $r_k$\,$<$\,$0.03$. This is consistent with the non-detection of a massive accreting disk around ROXs~12B by \cite{Bowler2017} and \cite{Wu2020}. The presence of circumsubstellar disks among wide-orbit companions is heterogeneous: while several systems show clear evidence of ongoing accretion or warm inner-disk emission \citep{Bowler2011, Bailey2013, Zhou2014}, others exhibit little to no detectable disk signatures \citep{Wu2017}. However, the retrieval of our injected test spectrum also failed to recover the included veiling ($r_k$\,=\,$0.2$), instead returning a lower metallicity (see Fig.~\ref{fig:summary_inj}). This demonstrates that modest veiling is challenging to detect at low S/N, so we cannot rule out its presence for ROXs~12B.

\subsection{Correlated noise}

Our results indicate a negligibly low degree of correlated noise for ROXs~12B, while for ROXs~12A it appears substantially higher. We find a GP amplitude $a$\,=\,8.9\,$\pm\,0.06$ and length-scale $\ell$\,=\,0.03\,$\pm\,0.01$\,nm\,$\approx$\,4.4\, pixels for the host star, while $a$\,=\,1.5\,$\pm\,0.02$ and $\ell$\,=\,0.01\,$\pm\,0.00$\,nm\,$\approx$\,1.4\,pixels for the companion. The large amplitude and length-scale for ROXs~12A is a direct result of our model being unable to reproduce several spectral features and is therefore not representative of the instrumentally correlated noise. This is further supported by the fact that we find barely any correlated noise for ROXs~12B, whose spectrum was extracted from the same data set.

\section{Discussion} \label{sec:discuss}

Determining an object's formation history is inherently challenging and requires multiple, often ambiguous, chemical and dynamical diagnostics. Objects forming within protoplanetary disks may develop supersolar C/O ratios ($>$\,0.59, \citealt{Asplund2021}) from accretion outside of the water snowline, as water ice freezes onto dust grains, enriching the C/O of the gas-phase \citep{Oberg2011, Brewer2017}. In contrast, C/O ratios near the solar value might indicate formation via cloud fragmentation or disk instability, where the composition reflects that of the parent molecular cloud \citep{Boss2011}. Our retrieved C/O ratio of 0.54\,$\pm$\,0.01 for ROXs~12B, which is close to solar, is consistent with formation through cloud fragmentation or disk instability. However, this diagnostic on its own is not definitive, particularly as we are unable to reliably compare it to its host star. The retrieved C/O ratio of $\approx$\,0.87 for ROXs~12A may not be robust as it misses the contribution from atomic oxygen, which is expected to be non-negligible at these temperatures. Similar C/O ratios have been found for other directly imaged companions in the SupJup survey, such as 0.59\,$\pm$\,0.01 for AB~Pic~b \citep{Gandhi2025}, C/O = 0.57\,$\pm$\,0.01 for YSES~1~b \citep{Zhang2024}, and 0.50\,$\pm$\,0.01 for GQ~Lup~B \citep{Picos2025}. Studies of other directly imaged planets also report near-solar C/O ratios, albeit with modest variations between systems \citep{Konopacky2013, Landman2024, Xuan2024}.

While the C/O ratio has long been suggested as a potential tracer of substellar formation pathways, the hypothesis that $^{12}$CO/$^{13}$CO might also serve as a formation tracer is fairly recent. The enrichment of $^{13}$C in the companion compared to the host star may indicate accretion of ices in the protoplanetary disk during its formation \citep{Zhang2021}, although the difference in isotopic ratios is only marginally significant within the 1-$\sigma$ uncertainties. Although the isotopic contrast between host and companion should be interpreted with caution as \texttt{MultiNest} may underestimate the uncertainties, this effect should be less pronounced for parameters that are not strongly affected by degeneracies, as supported by our injection-retrieval test. Since the isotopolog ratio is mainly constrained by the relative strengths of line features rather than their absolute depths, it depends less on parameters such as veiling and metallicity that scale the spectrum more globally and is therefore less strongly affected by degeneracies. Objects formed by core accretion are thought to inherit the $^{12}$C/$^{13}$C ratio of local disk solids, which is expected to vary throughout the disk due to isotope fractionation processes \citep{Visser2009, Yoshida2022}. This is in contrast to \cite{Xuan2024b} and \cite{Picos2025}, who find homogeneous isotopic compositions in the HIP~55507 and GQ~Lup systems, which may indicate that these systems were formed instead by gravitational collapse from the same parent cloud.

Interestingly, the possible core-accretion interpretation suggested by ROXs~12B's $^{12}$CO/$^{13}$CO ratio contrasts with the implications of its C/O ratio and the conclusions of \cite{Bowler2017}. They report a misalignment in the orbit of ROXs~12B with the spin axis of ROXs~12A by $49\,\substack{+20 \\ -32}^{\circ}$, which could indicate that it formed from a fragmenting molecular cloud core rather than in the disk of the star. Spin-orbit misalignments between widely-separated substellar companions and their hosts stars appear to be common \citep{Bowler2023}, in contrast to directly imaged planets in more compact systems \citep{Albrecht2013, Winn2017}. Hydrodynamic simulations also predict spin–orbit misalignments for binary systems that form via cloud fragmentation \citep{Bate2009, Offner2016}.

Together, these diagnostics paint an ambiguous picture of ROXs~12B's origin. Although both the near-solar C/O ratio and the spin–orbit misalignment favor a formation pathway involving gravitational collapse, the companion’s slightly lower $^{12}$C/$^{13}$C ratio compared to the host can instead be interpreted as pointing toward core accretion. Given these indications, the gravitational collapse scenario currently appears more plausible, but additional observational and modeling efforts will be required to possibly constrain the formation pathway of ROXs~12B and determine other complementary tracers.

However, the $^{12}$CO/$^{13}$CO ratio appears to be a more robust indicator of a system's age. Galactic chemical evolution models predict a gradual enrichment of the ISM in $^{13}$C over time, for instance via nova eruptions \citep{Romano2022}, a trend also seen in observations of nearby M dwarfs \citep{Picos2025_Mstars}. Given the young age of the ROXs~12 system, its $^{12}$CO/$^{13}$CO ratios are therefore expected to be close to the present ISM values, which we confirm in our analysis. Similar $^{12}$CO/$^{13}$CO ratios consistent with the ISM have been found in other young companions \citep{Zhang2024, Xuan2024, Picos2025} and young isolated brown dwarfs (\citealt{Picos2024}, \citealt{deRegt2026}).

\section{Conclusions}\label{sec:conclusion}

Our retrievals on high-resolution CRIRES+ spectroscopy of the ROXs~12 system reveal detailed insights into the atmospheres of the M0 host star and its L0 brown dwarf companion. We find evidence of $^{13}$CO and HF in both objects, as well as H$_2^{18}$O in the companion. Their $^{12}$CO/$^{13}$CO ratios of $77\,\substack{+10 \\ -7}$ and $55\,\substack{+10 \\ -7}$ for ROXs~12A and B are consistent with the ISM, which is expected due to their youth ($\sim$\,6\,Myrs). Furthermore, we obtain a C/O ratio of 0.54\,$\pm$\,0.01 for ROXs~12B, similar to other substellar companions, as well as a lower limit of H$_2^{16}$O/H$_2^{18}$O\,$\gtrapprox$\,300. The $P$--$T$ profile we retrieve for ROXs~12B exhibits shallower temperature gradients compared to the models, a phenomenon which has been observed in several other substellar companions in the literature. Although the formation mechanism of ROXs~12B remains ambiguous, ongoing and future efforts to measure the elemental and isotopic ratios in substellar companions and isolated brown dwarfs, such as those of the ESO SupJup Survey, will help shed more light on their physical and chemical properties, as well as help assess the potential of the $^{12}$CO/$^{13}$CO ratio as a tracer of formation history.

\begin{acknowledgements}
Support for this work was provided by NL-NWO Spinoza SPI.2022.004. Our work is based on observations collected at the European Organisation for Astronomical Research in the Southern Hemisphere under ESO programme 1110.C-4264. D.G.P., S.d.R. and I.S. acknowledge support from NWO grant OCENW.M.21.010. This research has made use of NASA’s Astrophysics Data System and the python packages NumPy \citep{Numpy2020}, SciPy \citep{Scipy2020}, Matplotlib \citep{Hunter2007}, petitRADTRANS \citep{Molliere2019}, PyAstronomy \citep{Czesla2019}), Astropy \citep{Astropy2022}, corner \citep{Foreman2016}.
\end{acknowledgements}

\bibliographystyle{aa}
\bibliography{article.bib}

\begin{thebibliography}{125}
\expandafter\ifx\csname natexlab\endcsname\relax\def\natexlab#1{#1}\fi

\bibitem[{{Albrecht} {et~al.}(2013){Albrecht}, {Winn}, {Marcy}, {Howard}, {Isaacson}, \& {Johnson}}]{Albrecht2013}
{Albrecht}, S., {Winn}, J.~N., {Marcy}, G.~W., {et~al.} 2013, \apj, 771, 11

\bibitem[{{Alcal{\'a}} {et~al.}(2021){Alcal{\'a}}, {Gangi}, {Biazzo}, {Antoniucci}, {Frasca}, {Giannini}, {Munari}, {Nisini}, {Harutyunyan}, {Manara}, \& {Vitali}}]{Alcala2021}
{Alcal{\'a}}, J.~M., {Gangi}, M., {Biazzo}, K., {et~al.} 2021, \aap, 652, A72

\bibitem[{{Allard} \& {Hauschildt}(1995)}]{Allard1995}
{Allard}, F. \& {Hauschildt}, P.~H. 1995, \apj, 445, 433

\bibitem[{{Allard} {et~al.}(2019){Allard}, {Spiegelman}, {Leininger}, \& {Molliere}}]{Allard2019}
{Allard}, N.~F., {Spiegelman}, F., {Leininger}, T., \& {Molliere}, P. 2019, \aap, 628, A120

\bibitem[{{Allers} \& {Liu}(2013)}]{Allers2013}
{Allers}, K.~N. \& {Liu}, M.~C. 2013, \apj, 772, 79

\bibitem[{{Antoniucci} {et~al.}(2017){Antoniucci}, {Nisini}, {Biazzo}, {Giannini}, {Lorenzetti}, {Sanna}, {Harutyunyan}, {Origlia}, \& {Oliva}}]{Antoniucci2017}
{Antoniucci}, S., {Nisini}, B., {Biazzo}, K., {et~al.} 2017, \aap, 606, A48

\bibitem[{{Asplund} {et~al.}(2021){Asplund}, {Amarsi}, \& {Grevesse}}]{Asplund2021}
{Asplund}, M., {Amarsi}, A.~M., \& {Grevesse}, N. 2021, \aap, 653, A141

\bibitem[{{Asplund} \& {Garc{\'\i}a P{\'e}rez}(2001)}]{Asplund2001}
{Asplund}, M. \& {Garc{\'\i}a P{\'e}rez}, A.~E. 2001, \aap, 372, 601

\bibitem[{{Astropy Collaboration} {et~al.}(2022){Astropy Collaboration}, {Price-Whelan}, {Lim}, {Earl}, {Starkman}, {Bradley}, {Shupe}, {Patil}, {Corrales}, {Brasseur}, {N{\"o}the}, {Donath}, {Tollerud}, {Morris}, {Ginsburg}, {Vaher}, {Weaver}, {Tocknell}, {Jamieson}, {van Kerkwijk}, {Robitaille}, {Merry}, {Bachetti}, {G{\"u}nther}, {Aldcroft}, {Alvarado-Montes}, {Archibald}, {B{\'o}di}, {Bapat}, {Barentsen}, {Baz{\'a}n}, {Biswas}, {Boquien}, {Burke}, {Cara}, {Cara}, {Conroy}, {Conseil}, {Craig}, {Cross}, {Cruz}, {D'Eugenio}, {Dencheva}, {Devillepoix}, {Dietrich}, {Eigenbrot}, {Erben}, {Ferreira}, {Foreman-Mackey}, {Fox}, {Freij}, {Garg}, {Geda}, {Glattly}, {Gondhalekar}, {Gordon}, {Grant}, {Greenfield}, {Groener}, {Guest}, {Gurovich}, {Handberg}, {Hart}, {Hatfield-Dodds}, {Homeier}, {Hosseinzadeh}, {Jenness}, {Jones}, {Joseph}, {Kalmbach}, {Karamehmetoglu}, {Ka{\l}uszy{\'n}ski}, {Kelley}, {Kern}, {Kerzendorf}, {Koch}, {Kulumani}, {Lee}, {Ly}, {Ma}, {MacBride}, {Maljaars}, {Muna}, {Murphy}, {Norman},
  {O'Steen}, {Oman}, {Pacifici}, {Pascual}, {Pascual-Granado}, {Patil}, {Perren}, {Pickering}, {Rastogi}, {Roulston}, {Ryan}, {Rykoff}, {Sabater}, {Sakurikar}, {Salgado}, {Sanghi}, {Saunders}, {Savchenko}, {Schwardt}, {Seifert-Eckert}, {Shih}, {Jain}, {Shukla}, {Sick}, {Simpson}, {Singanamalla}, {Singer}, {Singhal}, {Sinha}, {Sip{\H{o}}cz}, {Spitler}, {Stansby}, {Streicher}, {{\v{S}}umak}, {Swinbank}, {Taranu}, {Tewary}, {Tremblay}, {de Val-Borro}, {Van Kooten}, {Vasovi{\'c}}, {Verma}, {de Miranda Cardoso}, {Williams}, {Wilson}, {Winkel}, {Wood-Vasey}, {Xue}, {Yoachim}, {Zhang}, {Zonca}, \& {Astropy Project Contributors}}]{Astropy2022}
{Astropy Collaboration}, {Price-Whelan}, A.~M., {Lim}, P.~L., {et~al.} 2022, \apj, 935, 167

\bibitem[{{Bailey} {et~al.}(2013){Bailey}, {Hinz}, {Currie}, {Su}, {Esposito}, {Hill}, {Hoffmann}, {Jones}, {Kim}, {Leisenring}, {Meyer}, {Murray-Clay}, {Nelson}, {Pinna}, {Puglisi}, {Rieke}, {Rodigas}, {Skemer}, {Skrutskie}, {Vaitheeswaran}, \& {Wilson}}]{Bailey2013}
{Bailey}, V., {Hinz}, P.~M., {Currie}, T., {et~al.} 2013, \apj, 767, 31

\bibitem[{{Baraffe} {et~al.}(2015){Baraffe}, {Homeier}, {Allard}, \& {Chabrier}}]{Baraffe2015}
{Baraffe}, I., {Homeier}, D., {Allard}, F., \& {Chabrier}, G. 2015, \aap, 577, A42

\bibitem[{{Bate}(2009)}]{Bate2009}
{Bate}, M.~R. 2009, \mnras, 392, 590

\bibitem[{{Bate} {et~al.}(2002){Bate}, {Bonnell}, \& {Bromm}}]{Bate2002}
{Bate}, M.~R., {Bonnell}, I.~A., \& {Bromm}, V. 2002, \mnras, 332, L65

\bibitem[{{Bergin} {et~al.}(2024){Bergin}, {Bosman}, {Teague}, {Calahan}, {Willacy}, {Cleeves}, {Schwarz}, {Zhang}, \& {Bruderer}}]{Bergin2024}
{Bergin}, E.~A., {Bosman}, A., {Teague}, R., {et~al.} 2024, \apj, 965, 147

\bibitem[{{Borysow} {et~al.}(1988){Borysow}, {Frommhold}, \& {Birnbaum}}]{Borysow1988}
{Borysow}, J., {Frommhold}, L., \& {Birnbaum}, G. 1988, \apj, 326, 509

\bibitem[{{Boss}(1997)}]{Boss1997}
{Boss}, A.~P. 1997, Science, 276, 1836

\bibitem[{{Boss}(2001)}]{Boss2001}
{Boss}, A.~P. 2001, \apjl, 551, L167

\bibitem[{{Boss}(2006)}]{Boss2006}
{Boss}, A.~P. 2006, \apjl, 637, L137

\bibitem[{{Boss}(2011)}]{Boss2011}
{Boss}, A.~P. 2011, \apj, 731, 74

\bibitem[{{Bouvier} \& {Appenzeller}(1992)}]{Bouvier1992}
{Bouvier}, J. \& {Appenzeller}, I. 1992, \aaps, 92, 481

\bibitem[{{Bowler} {et~al.}(2017){Bowler}, {Kraus}, {Bryan}, {Knutson}, {Brogi}, {Rizzuto}, {Mace}, {Vanderburg}, {Liu}, {Hillenbrand}, \& {Cieza}}]{Bowler2017}
{Bowler}, B.~P., {Kraus}, A.~L., {Bryan}, M.~L., {et~al.} 2017, \aj, 154, 165

\bibitem[{{Bowler} {et~al.}(2011){Bowler}, {Liu}, {Kraus}, {Mann}, \& {Ireland}}]{Bowler2011}
{Bowler}, B.~P., {Liu}, M.~C., {Kraus}, A.~L., {Mann}, A.~W., \& {Ireland}, M.~J. 2011, \apj, 743, 148

\bibitem[{{Bowler} {et~al.}(2023){Bowler}, {Tran}, {Zhang}, {Morgan}, {Ashok}, {Blunt}, {Bryan}, {Evans}, {Franson}, {Huber}, {Nagpal}, {Wu}, \& {Zhou}}]{Bowler2023}
{Bowler}, B.~P., {Tran}, Q.~H., {Zhang}, Z., {et~al.} 2023, \aj, 165, 164

\bibitem[{{Brewer} {et~al.}(2017){Brewer}, {Fischer}, \& {Madhusudhan}}]{Brewer2017}
{Brewer}, J.~M., {Fischer}, D.~A., \& {Madhusudhan}, N. 2017, \aj, 153, 83

\bibitem[{{Buchner}(2016)}]{Buchner2016}
{Buchner}, J. 2016, Statistics and Computing, 26, 383

\bibitem[{{Buchner} {et~al.}(2014){Buchner}, {Georgakakis}, {Nandra}, {Hsu}, {Rangel}, {Brightman}, {Merloni}, {Salvato}, {Donley}, \& {Kocevski}}]{Buchner2014}
{Buchner}, J., {Georgakakis}, A., {Nandra}, K., {et~al.} 2014, \aap, 564, A125

\bibitem[{{Burningham} {et~al.}(2021){Burningham}, {Faherty}, {Gonzales}, {Marley}, {Visscher}, {Lupu}, {Gaarn}, {Fabienne Bieger}, {Freedman}, \& {Saumon}}]{Burningham2021}
{Burningham}, B., {Faherty}, J.~K., {Gonzales}, E.~C., {et~al.} 2021, \mnras, 506, 1944

\bibitem[{{Burningham} {et~al.}(2017){Burningham}, {Marley}, {Line}, {Lupu}, {Visscher}, {Morley}, {Saumon}, \& {Freedman}}]{Burningham2017}
{Burningham}, B., {Marley}, M.~S., {Line}, M.~R., {et~al.} 2017, \mnras, 470, 1177

\bibitem[{{Castelli} \& {Kurucz}(2003)}]{Kurucz2003}
{Castelli}, F. \& {Kurucz}, R.~L. 2003, in IAU Symposium, Vol. 210, Modelling of Stellar Atmospheres, ed. N.~{Piskunov}, W.~W. {Weiss}, \& D.~F. {Gray}, A20

\bibitem[{{Chan} \& {Dalgarno}(1965)}]{Chan1965}
{Chan}, Y.~M. \& {Dalgarno}, A. 1965, Proceedings of the Physical Society, 85, 227

\bibitem[{{Coxon} \& {Hajigeorgiou}(2015)}]{Coxon2015}
{Coxon}, J.~A. \& {Hajigeorgiou}, P.~G. 2015, \jqsrt, 151, 133

\bibitem[{{Cristofari} {et~al.}(2022){Cristofari}, {Donati}, {Masseron}, {Fouqu{\'e}}, {Moutou}, {Carmona}, {Artigau}, {Martioli}, {H{\'e}brard}, {Gaidos}, {Delfosse}, \& {SLS consortium}}]{Cristofari2022}
{Cristofari}, P.~I., {Donati}, J.-F., {Masseron}, T., {et~al.} 2022, \mnras, 516, 3802

\bibitem[{{Cugno} {et~al.}(2024){Cugno}, {Patapis}, {Banzatti}, {Meyer}, {Dannert}, {Stolker}, {MacDonald}, \& {Pontoppidan}}]{Cugno2024}
{Cugno}, G., {Patapis}, P., {Banzatti}, A., {et~al.} 2024, \apjl, 966, L21

\bibitem[{{Czesla} {et~al.}(2019){Czesla}, {Schr{\"o}ter}, {Schneider}, {Huber}, {Pfeifer}, {Andreasen}, \& {Zechmeister}}]{Czesla2019}
{Czesla}, S., {Schr{\"o}ter}, S., {Schneider}, C.~P., {et~al.} 2019, {PyA: Python astronomy-related packages}, Astrophysics Source Code Library, record ascl:1906.010

\bibitem[{{Dalgarno} \& {Williams}(1962)}]{Dalgarno1962}
{Dalgarno}, A. \& {Williams}, D.~A. 1962, \apj, 136, 690

\bibitem[{{de Regt} {et~al.}(2024){de Regt}, {Gandhi}, {Snellen}, {Zhang}, {Ginski}, {Gonz{\'a}lez Picos}, {Kesseli}, {Landman}, {Molli{\`e}re}, {Nasedkin}, {S{\'a}nchez-L{\'o}pez}, \& {Stolker}}]{deRegt2024}
{de Regt}, S., {Gandhi}, S., {Snellen}, I.~A.~G., {et~al.} 2024, \aap, 688, A116

\bibitem[{{de Regt} {et~al.}(2026){de Regt}, {Snellen}, {Gonz{\'a}lez Picos}, {Gandhi}, {Grasser}, {Kesseli}, {Landman}, {Molli{\`e}re}, {Nasedkin}, {Stolker}, \& {Zhang}}]{deRegt2026}
{de Regt}, S., {Snellen}, I.~A.~G., {Gonz{\'a}lez Picos}, D., {et~al.} 2026, \aap, 707, A210

\bibitem[{{Deacon} {et~al.}(2016){Deacon}, {Schlieder}, \& {Murphy}}]{Deacon2016}
{Deacon}, N.~R., {Schlieder}, J.~E., \& {Murphy}, S.~J. 2016, \mnras, 457, 3191

\bibitem[{{Dittmann}(2024)}]{Dittmann2024}
{Dittmann}, A. 2024, The Open Journal of Astrophysics, 7, 79

\bibitem[{{Dorn} {et~al.}(2014){Dorn}, {Anglada-Escude}, {Baade}, {Bristow}, {Follert}, {Gojak}, {Grunhut}, {Hatzes}, {Heiter}, {Hilker}, {Ives}, {Jung}, {K{\"a}ufl}, {Kerber}, {Klein}, {Lizon}, {Lockhart}, {L{\"o}winger}, {Marquart}, {Oliva}, {Origlia}, {Pasquini}, {Paufique}, {Piskunov}, {Pozna}, {Reiners}, {Smette}, {Smoker}, {Seemann}, {Stempels}, \& {Valenti}}]{Dorn2014}
{Dorn}, R.~J., {Anglada-Escude}, G., {Baade}, D., {et~al.} 2014, The Messenger, 156, 7

\bibitem[{{Feroz} {et~al.}(2009){Feroz}, {Hobson}, \& {Bridges}}]{Feroz2009}
{Feroz}, F., {Hobson}, M.~P., \& {Bridges}, M. 2009, \mnras, 398, 1601

\bibitem[{{Feroz} {et~al.}(2019){Feroz}, {Hobson}, {Cameron}, \& {Pettitt}}]{Feroz2019}
{Feroz}, F., {Hobson}, M.~P., {Cameron}, E., \& {Pettitt}, A.~N. 2019, The Open Journal of Astrophysics, 2, 10

\bibitem[{{Filippazzo} {et~al.}(2015){Filippazzo}, {Rice}, {Faherty}, {Cruz}, {Van Gordon}, \& {Looper}}]{Filippazzo2015}
{Filippazzo}, J.~C., {Rice}, E.~L., {Faherty}, J., {et~al.} 2015, \apj, 810, 158

\bibitem[{{Fischer} {et~al.}(2011){Fischer}, {Edwards}, {Hillenbrand}, \& {Kwan}}]{Fischer2011}
{Fischer}, W., {Edwards}, S., {Hillenbrand}, L., \& {Kwan}, J. 2011, \apj, 730, 73

\bibitem[{{Follert} {et~al.}(2014){Follert}, {Dorn}, {Oliva}, {Lizon}, {Hatzes}, {Piskunov}, {Reiners}, {Seemann}, {Stempels}, {Heiter}, {Marquart}, {Lockhart}, {Anglada-Escude}, {L{\"o}winger}, {Baade}, {Grunhut}, {Bristow}, {Klein}, {Jung}, {Ives}, {Kerber}, {Pozna}, {Paufique}, {Kaeufl}, {Origlia}, {Valenti}, {Gojak}, {Hilker}, {Pasquini}, {Smette}, \& {Smoker}}]{Follert2014}
{Follert}, R., {Dorn}, R.~J., {Oliva}, E., {et~al.} 2014, in Instrumentation, 914719

\bibitem[{{Foreman-Mackey}(2016)}]{Foreman2016}
{Foreman-Mackey}, D. 2016, The Journal of Open Source Software, 1, 24

\bibitem[{{Gaia Collaboration}(2020)}]{Gaia2020}
{Gaia Collaboration}. 2020, VizieR Online Data Catalog, I/350

\bibitem[{{Gandhi} {et~al.}(2025){Gandhi}, {de Regt}, {Snellen}, {Palma-Bifani}, {Abdoulwahab}, {Chauvin}, {Gonz{\'a}lez Picos}, {Zhang}, {Landman}, {Stolker}, {Kesseli}, {Mulder}, {Chomez}, {Lagrange}, \& {Zurlo}}]{Gandhi2025}
{Gandhi}, S., {de Regt}, S., {Snellen}, I., {et~al.} 2025, \mnras, 537, 134

\bibitem[{{Gonz{\'a}lez Picos} {et~al.}(2025{\natexlab{a}}){Gonz{\'a}lez Picos}, {Snellen}, \& {de Regt}}]{Picos2025_Mstars}
{Gonz{\'a}lez Picos}, D., {Snellen}, I., \& {de Regt}, S. 2025{\natexlab{a}}, Nature Astronomy, 9, 1692

\bibitem[{{Gonz{\'a}lez Picos} {et~al.}(2024){Gonz{\'a}lez Picos}, {Snellen}, {de Regt}, {Landman}, {Zhang}, {Gandhi}, {Ginski}, {Kesseli}, {Molli{\`e}re}, \& {Stolker}}]{Picos2024}
{Gonz{\'a}lez Picos}, D., {Snellen}, I.~A.~G., {de Regt}, S., {et~al.} 2024, \aap, 689, A212

\bibitem[{{Gonz{\'a}lez Picos} {et~al.}(2025{\natexlab{b}}){Gonz{\'a}lez Picos}, {Snellen}, {de Regt}, {Landman}, {Zhang}, {Gandhi}, \& {S{\'a}nchez-L{\'o}pez}}]{Picos2025}
{Gonz{\'a}lez Picos}, D., {Snellen}, I.~A.~G., {de Regt}, S., {et~al.} 2025{\natexlab{b}}, \aap, 693, A298

\bibitem[{{Grasser} {et~al.}(2025){Grasser}, {Snellen}, {de Regt}, {Gonz{\'a}lez Picos}, {Zhang}, {Stolker}, {Gandhi}, {Nasedkin}, {Landman}, {Kesseli}, \& {Mulder}}]{Grasser2025}
{Grasser}, N., {Snellen}, I.~A.~G., {de Regt}, S., {et~al.} 2025, \aap, 698, A252

\bibitem[{{Gray}(2008)}]{Gray2008}
{Gray}, D.~F. 2008, {The Observation and Analysis of Stellar Photospheres} (Cambridge University Press)

\bibitem[{Harris {et~al.}(2020)Harris, Millman, van~der Walt, Gommers, Virtanen, Cournapeau, Wieser, Taylor, Berg, Smith, Kern, Picus, Hoyer, van Kerkwijk, Brett, Haldane, Fernández~del Río, Wiebe, Peterson, Gérard-Marchant, Sheppard, Reddy, Weckesser, Abbasi, Gohlke, \& Oliphant}]{Numpy2020}
Harris, C.~R., Millman, K.~J., van~der Walt, S.~J., {et~al.} 2020, Nature, 585, 357–362

\bibitem[{{Holmberg} \& {Madhusudhan}(2022)}]{HolmbergMadu2022}
{Holmberg}, M. \& {Madhusudhan}, N. 2022, \aj, 164, 79

\bibitem[{{Horne}(1986)}]{Horne1986}
{Horne}, K. 1986, \pasp, 98, 609

\bibitem[{{Hunter}(2007)}]{Hunter2007}
{Hunter}, J.~D. 2007, Computing in Science and Engineering, 9, 90

\bibitem[{{Inglis} {et~al.}(2024){Inglis}, {Wallack}, {Xuan}, {Knutson}, {Chachan}, {Bryan}, {Bowler}, {Iyer}, {Kataria}, \& {Benneke}}]{Inglis2024}
{Inglis}, J., {Wallack}, N.~L., {Xuan}, J.~W., {et~al.} 2024, \aj, 167, 218

\bibitem[{{Iyer} {et~al.}(2023){Iyer}, {Line}, {Muirhead}, {Fortney}, \& {Gharib-Nezhad}}]{Iyer2023_SPHINX}
{Iyer}, A.~R., {Line}, M.~R., {Muirhead}, P.~S., {Fortney}, J.~J., \& {Gharib-Nezhad}, E. 2023, \apj, 944, 41

\bibitem[{{Kaeufl} {et~al.}(2004){Kaeufl}, {Ballester}, {Biereichel}, {Delabre}, {Donaldson}, {Dorn}, {Fedrigo}, {Finger}, {Fischer}, {Franza}, {Gojak}, {Huster}, {Jung}, {Lizon}, {Mehrgan}, {Meyer}, {Moorwood}, {Pirard}, {Paufique}, {Pozna}, {Siebenmorgen}, {Silber}, {Stegmeier}, \& {Wegerer}}]{Kaeufl2004}
{Kaeufl}, H.-U., {Ballester}, P., {Biereichel}, P., {et~al.} 2004, in Society of Photo-Optical Instrumentation Engineers (SPIE) Conference Series, Vol. 5492, Ground-based Instrumentation for Astronomy, ed. A.~F.~M. {Moorwood} \& M.~{Iye}, 1218--1227

\bibitem[{{Kawahara} {et~al.}(2022){Kawahara}, {Kawashima}, {Masuda}, {Crossfield}, {Pannier}, \& {van den Bekerom}}]{Kawahara2022}
{Kawahara}, H., {Kawashima}, Y., {Masuda}, K., {et~al.} 2022, \apjs, 258, 31

\bibitem[{{Kitzmann} {et~al.}(2020){Kitzmann}, {Heng}, {Oreshenko}, {Grimm}, {Apai}, {Bowler}, {Burgasser}, \& {Marley}}]{Kitzmann2020}
{Kitzmann}, D., {Heng}, K., {Oreshenko}, M., {et~al.} 2020, \apj, 890, 174

\bibitem[{{Kitzmann} {et~al.}(2024){Kitzmann}, {Stock}, \& {Patzer}}]{Kitzmann2024}
{Kitzmann}, D., {Stock}, J.~W., \& {Patzer}, A. B.~C. 2024, \mnras, 527, 7263

\bibitem[{{Konopacky} {et~al.}(2013){Konopacky}, {Barman}, {Macintosh}, \& {Marois}}]{Konopacky2013}
{Konopacky}, Q.~M., {Barman}, T.~S., {Macintosh}, B.~A., \& {Marois}, C. 2013, Science, 339, 1398

\bibitem[{{Kratter} \& {Lodato}(2016)}]{Kratter2016}
{Kratter}, K. \& {Lodato}, G. 2016, \araa, 54, 271

\bibitem[{{Kraus} {et~al.}(2014){Kraus}, {Ireland}, {Cieza}, {Hinkley}, {Dupuy}, {Bowler}, \& {Liu}}]{Kraus2014}
{Kraus}, A.~L., {Ireland}, M.~J., {Cieza}, L.~A., {et~al.} 2014, \apj, 781, 20

\bibitem[{{Kuzuhara} {et~al.}(2022){Kuzuhara}, {Currie}, {Takarada}, {Brandt}, {Sato}, {Uyama}, {Janson}, {Chilcote}, {Tobin}, {Lawson}, {Hori}, {Guyon}, {Groff}, {Lozi}, {Vievard}, {Sahoo}, {Deo}, {Jovanovic}, {Ahn}, {Martinache}, {Skaf}, {Akiyama}, {Norris}, {Bonnefoy}, {He{\l}miniak}, {Kudo}, {McElwain}, {Samland}, {Wagner}, {Wisniewski}, {Knapp}, {Kwon}, {Nishikawa}, {Serabyn}, {Hayashi}, \& {Tamura}}]{Kuzuhara2022}
{Kuzuhara}, M., {Currie}, T., {Takarada}, T., {et~al.} 2022, \apjl, 934, L18

\bibitem[{{Landman} {et~al.}(2024){Landman}, {Stolker}, {Snellen}, {Costes}, {de Regt}, {Zhang}, {Gandhi}, {Molliere}, {Kesseli}, {Vigan}, \& {Sanchez-L{\'o}pez}}]{Landman2024}
{Landman}, R., {Stolker}, T., {Snellen}, I.~A.~G., {et~al.} 2024, \aap, 682, A48

\bibitem[{{Lemos} {et~al.}(2023){Lemos}, {Weaverdyck}, {Rollins}, {Muir}, {Fert{\'e}}, {Liddle}, {Campos}, {Huterer}, {Raveri}, {Zuntz}, {Di Valentino}, {Fang}, {Hartley}, {Aguena}, {Allam}, {Annis}, {Bertin}, {Bocquet}, {Brooks}, {Burke}, {Carnero Rosell}, {Carrasco Kind}, {Carretero}, {Castander}, {Choi}, {Costanzi}, {Crocce}, {da Costa}, {Pereira}, {Dietrich}, {Everett}, {Ferrero}, {Frieman}, {Garc{\'\i}a-Bellido}, {Gatti}, {Gaztanaga}, {Gerdes}, {Gruen}, {Gruendl}, {Gschwend}, {Gutierrez}, {Hinton}, {Hollowood}, {Honscheid}, {James}, {Kuehn}, {Kuropatkin}, {Lima}, {March}, {Melchior}, {Menanteau}, {Miquel}, {Morgan}, {Palmese}, {Paz-Chinch{\'o}n}, {Pieres}, {Malag{\'o}n}, {Porredon}, {Sanchez}, {Scarpine}, {Schubnell}, {Serrano}, {Sevilla-Noarbe}, {Smith}, {Suchyta}, {Swanson}, {Tarle}, {Thomas}, {To}, {Varga}, {Weller}, \& {DES Collaboration}}]{Lemos2023}
{Lemos}, P., {Weaverdyck}, N., {Rollins}, R.~P., {et~al.} 2023, \mnras, 521, 1184

\bibitem[{{Li} {et~al.}(2013){Li}, {Gordon}, {Le Roy}, {Hajigeorgiou}, {Coxon}, {Bernath}, \& {Rothman}}]{Li2013}
{Li}, G., {Gordon}, I.~E., {Le Roy}, R.~J., {et~al.} 2013, \jqsrt, 121, 78

\bibitem[{{Li} {et~al.}(2015){Li}, {Gordon}, {Rothman}, {Tan}, {Hu}, {Kassi}, {Campargue}, \& {Medvedev}}]{Li2015}
{Li}, G., {Gordon}, I.~E., {Rothman}, L.~S., {et~al.} 2015, \apjs, 216, 15

\bibitem[{{Line} {et~al.}(2021){Line}, {Brogi}, {Bean}, {Gandhi}, {Zalesky}, {Parmentier}, {Smith}, {Mace}, {Mansfield}, {Kempton}, {Fortney}, {Shkolnik}, {Patience}, {Rauscher}, {D{\'e}sert}, \& {Wardenier}}]{Line2021}
{Line}, M.~R., {Brogi}, M., {Bean}, J.~L., {et~al.} 2021, \nat, 598, 580

\bibitem[{{Lueber} {et~al.}(2022){Lueber}, {Kitzmann}, {Bowler}, {Burgasser}, \& {Heng}}]{Lueber2022}
{Lueber}, A., {Kitzmann}, D., {Bowler}, B.~P., {Burgasser}, A.~J., \& {Heng}, K. 2022, \apj, 930, 136

\bibitem[{{Lyons} {et~al.}(2018){Lyons}, {Gharib-Nezhad}, \& {Ayres}}]{Lyons2018}
{Lyons}, J.~R., {Gharib-Nezhad}, E., \& {Ayres}, T.~R. 2018, Nature Communications, 9, 908

\bibitem[{{Marley} {et~al.}(2021){Marley}, {Saumon}, {Visscher}, {Lupu}, {Freedman}, {Morley}, {Fortney}, {Seay}, {Smith}, {Teal}, \& {Wang}}]{Marley2021_SonoraBobcat}
{Marley}, M.~S., {Saumon}, D., {Visscher}, C., {et~al.} 2021, \apj, 920, 85

\bibitem[{{Milam} {et~al.}(2005){Milam}, {Savage}, {Brewster}, {Ziurys}, \& {Wyckoff}}]{Milam2005}
{Milam}, S.~N., {Savage}, C., {Brewster}, M.~A., {Ziurys}, L.~M., \& {Wyckoff}, S. 2005, \apj, 634, 1126

\bibitem[{{Molli{\`e}re} \& {Snellen}(2019)}]{MolliereSnellen2019}
{Molli{\`e}re}, P. \& {Snellen}, I.~A.~G. 2019, \aap, 622, A139

\bibitem[{{Molli{\`e}re} {et~al.}(2020){Molli{\`e}re}, {Stolker}, {Lacour}, {Otten}, {Shangguan}, {Charnay}, {Molyarova}, {Nowak}, {Henning}, {Marleau}, {Semenov}, {van Dishoeck}, {Eisenhauer}, {Garcia}, {Garcia Lopez}, {Girard}, {Greenbaum}, {Hinkley}, {Kervella}, {Kreidberg}, {Maire}, {Nasedkin}, {Pueyo}, {Snellen}, {Vigan}, {Wang}, {de Zeeuw}, \& {Zurlo}}]{Molliere2020}
{Molli{\`e}re}, P., {Stolker}, T., {Lacour}, S., {et~al.} 2020, \aap, 640, A131

\bibitem[{{Molli{\`e}re} {et~al.}(2019){Molli{\`e}re}, {Wardenier}, {van Boekel}, {Henning}, {Molaverdikhani}, \& {Snellen}}]{Molliere2019}
{Molli{\`e}re}, P., {Wardenier}, J.~P., {van Boekel}, R., {et~al.} 2019, \aap, 627, A67

\bibitem[{{Montmerle} {et~al.}(1983){Montmerle}, {Koch-Miramond}, {Falgarone}, \& {Grindlay}}]{Montmerle1983}
{Montmerle}, T., {Koch-Miramond}, L., {Falgarone}, E., \& {Grindlay}, J.~E. 1983, \apj, 269, 182

\bibitem[{{Morley} {et~al.}(2019){Morley}, {Skemer}, {Miles}, {Line}, {Lopez}, {Brogi}, {Freedman}, \& {Marley}}]{Morley2019}
{Morley}, C.~V., {Skemer}, A.~J., {Miles}, B.~E., {et~al.} 2019, \apjl, 882, L29

\bibitem[{{Mulder} {et~al.}(2025){Mulder}, {de Regt}, {Landman}, {Picos}, {Snellen}, {Zhang}, {Gandhi}, {Ginski}, {Kesseli}, {Nasedkin}, \& {Stolker}}]{Mulder2025}
{Mulder}, W., {de Regt}, S., {Landman}, R., {et~al.} 2025, \aap, 694, A164

\bibitem[{{Muzerolle} {et~al.}(2003){Muzerolle}, {Calvet}, {Hartmann}, \& {D'Alessio}}]{Muzerolle2003}
{Muzerolle}, J., {Calvet}, N., {Hartmann}, L., \& {D'Alessio}, P. 2003, \apjl, 597, L149

\bibitem[{{{\"O}berg} \& {Bergin}(2021)}]{Oberg2021}
{{\"O}berg}, K.~I. \& {Bergin}, E.~A. 2021, \physrep, 893, 1

\bibitem[{{{\"O}berg} {et~al.}(2011){{\"O}berg}, {Murray-Clay}, \& {Bergin}}]{Oberg2011}
{{\"O}berg}, K.~I., {Murray-Clay}, R., \& {Bergin}, E.~A. 2011, \apjl, 743, L16

\bibitem[{{Offner} {et~al.}(2016){Offner}, {Dunham}, {Lee}, {Arce}, \& {Fielding}}]{Offner2016}
{Offner}, S. S.~R., {Dunham}, M.~M., {Lee}, K.~I., {Arce}, H.~G., \& {Fielding}, D.~B. 2016, \apjl, 827, L11

\bibitem[{{Olander} {et~al.}(2021){Olander}, {Heiter}, \& {Kochukhov}}]{Olander2021}
{Olander}, T., {Heiter}, U., \& {Kochukhov}, O. 2021, \aap, 649, A103

\bibitem[{{Pacetti} {et~al.}(2022){Pacetti}, {Turrini}, {Schisano}, {Molinari}, {Fonte}, {Politi}, {Hennebelle}, {Klessen}, {Testi}, \& {Lebreuilly}}]{Pacetti2022}
{Pacetti}, E., {Turrini}, D., {Schisano}, E., {et~al.} 2022, \apj, 937, 36

\bibitem[{{Phillips} {et~al.}(2020){Phillips}, {Tremblin}, {Baraffe}, {Chabrier}, {Allard}, {Spiegelman}, {Goyal}, {Drummond}, \& {H{\'e}brard}}]{Phillips2020}
{Phillips}, M.~W., {Tremblin}, P., {Baraffe}, I., {et~al.} 2020, \aap, 637, A38

\bibitem[{{Polyansky} {et~al.}(2017){Polyansky}, {Kyuberis}, {Lodi}, {Tennyson}, {Yurchenko}, {Ovsyannikov}, \& {Zobov}}]{Polyansky2017}
{Polyansky}, O.~L., {Kyuberis}, A.~A., {Lodi}, L., {et~al.} 2017, \mnras, 466, 1363

\bibitem[{{Polyansky} {et~al.}(2018){Polyansky}, {Kyuberis}, {Zobov}, {Tennyson}, {Yurchenko}, \& {Lodi}}]{Polyansky2018}
{Polyansky}, O.~L., {Kyuberis}, A.~A., {Zobov}, N.~F., {et~al.} 2018, \mnras, 480, 2597

\bibitem[{{Rajpurohit} {et~al.}(2018){Rajpurohit}, {Allard}, {Rajpurohit}, {Sharma}, {Teixeira}, {Mousis}, \& {Rajpurohit}}]{Rajpurohit2018}
{Rajpurohit}, A.~S., {Allard}, F., {Rajpurohit}, S., {et~al.} 2018, \aap, 620, A180

\bibitem[{{Ratzka} {et~al.}(2005){Ratzka}, {K{\"o}hler}, \& {Leinert}}]{Ratzka2005}
{Ratzka}, T., {K{\"o}hler}, R., \& {Leinert}, C. 2005, \aap, 437, 611

\bibitem[{{Reggiani} {et~al.}(2019){Reggiani}, {Amarsi}, {Lind}, {Barklem}, {Zatsarinny}, {Bartschat}, {Fursa}, {Bray}, {Spina}, \& {Mel{\'e}ndez}}]{Reggiani2019}
{Reggiani}, H., {Amarsi}, A.~M., {Lind}, K., {et~al.} 2019, \aap, 627, A177

\bibitem[{{Romano}(2022)}]{Romano2022}
{Romano}, D. 2022, \aapr, 30, 7

\bibitem[{{Royer} {et~al.}(2024){Royer}, {Merle}, {Dsilva}, {Sekaran}, {Van Winckel}, {Fr{\'e}mat}, {Van der Swaelmen}, {Gebruers}, {Tkachenko}, {Laverick}, {Dirickx}, {Raskin}, {Hensberge}, {Abdul-Masih}, {Acke}, {Alonso}, {Bandhu Mahato}, {Beck}, {Behara}, {Bloemen}, {Buysschaert}, {Cox}, {Debosscher}, {De Cat}, {Degroote}, {De Nutte}, {De Smedt}, {de Vries}, {Dumortier}, {Escorza}, {Exter}, {Goriely}, {Gorlova}, {Hillen}, {Homan}, {Jorissen}, {Kamath}, {Karjalainen}, {Karjalainen}, {Lampens}, {Lobel}, {Lombaert}, {Marcos-Arenal}, {Menu}, {Merges}, {Moravveji}, {Nemeth}, {Neyskens}, {Ostensen}, {P{\'a}pics}, {Perez}, {Prins}, {Royer}, {Samadi-Ghadim}, {Sana}, {Sans Fuentes}, {Scaringi}, {Schmid}, {Siess}, {Siopis}, {Smolders}, {S{\'o}dor}, {Thoul}, {Triana}, {Vandenbussche}, {Van de Sande}, {Van De Steene}, {Van Eck}, {van Hoof}, {Van Marle}, {Van Reeth}, {Vermeylen}, {Volpi}, {Vos}, \& {Waelkens}}]{Royer2024}
{Royer}, P., {Merle}, T., {Dsilva}, K., {et~al.} 2024, \aap, 681, A107

\bibitem[{{Ruffio} {et~al.}(2019){Ruffio}, {Macintosh}, {Konopacky}, {Barman}, {De Rosa}, {Wang}, {Wilcomb}, {Czekala}, \& {Marois}}]{Ruffio2019}
{Ruffio}, J.-B., {Macintosh}, B., {Konopacky}, Q.~M., {et~al.} 2019, \aj, 158, 200

\bibitem[{{Santos} {et~al.}(2008){Santos}, {Melo}, {James}, {Gameiro}, {Bouvier}, \& {Gomes}}]{Santos2008}
{Santos}, N.~C., {Melo}, C., {James}, D.~J., {et~al.} 2008, \aap, 480, 889

\bibitem[{{Smette} {et~al.}(2015){Smette}, {Sana}, {Noll}, {Horst}, {Kausch}, {Kimeswenger}, {Barden}, {Szyszka}, {Jones}, {Gallenne}, {Vinther}, {Ballester}, \& {Taylor}}]{Smette2015}
{Smette}, A., {Sana}, H., {Noll}, S., {et~al.} 2015, \aap, 576, A77

\bibitem[{{Sokal} {et~al.}(2018){Sokal}, {Deen}, {Mace}, {Lee}, {Oh}, {Kim}, {Kidder}, \& {Jaffe}}]{Sokal2018}
{Sokal}, K.~R., {Deen}, C.~P., {Mace}, G.~N., {et~al.} 2018, \apj, 853, 120

\bibitem[{{Somogyi} {et~al.}(2021){Somogyi}, {Yurchenko}, \& {Yachmenev}}]{Somogyi2021}
{Somogyi}, W., {Yurchenko}, S.~N., \& {Yachmenev}, A. 2021, \jcp, 155, 214303

\bibitem[{{Sousa} {et~al.}(2023){Sousa}, {Bouvier}, {Alencar}, {Donati}, {Dougados}, {Alecian}, {Carmona}, {Rebull}, {Cook}, {Artigau}, {Fouqu{\'e}}, \& {Doyon}}]{Sousa2023}
{Sousa}, A.~P., {Bouvier}, J., {Alencar}, S.~H.~P., {et~al.} 2023, \aap, 670, A142

\bibitem[{{Stolker} \& {Landman}(2023)}]{StolkerLandman2023}
{Stolker}, T. \& {Landman}, R. 2023, {pycrires: Data reduction pipeline for VLT/CRIRES+}, Astrophysics Source Code Library, record ascl:2307.040

\bibitem[{{Sullivan} {et~al.}(2019){Sullivan}, {Wilking}, {Greene}, {Lisalda}, {Gibb}, \& {Ejeta}}]{Sullivan2019}
{Sullivan}, T., {Wilking}, B.~A., {Greene}, T.~P., {et~al.} 2019, \aj, 158, 41

\bibitem[{{Swastik} {et~al.}(2021){Swastik}, {Banyal}, {Narang}, {Manoj}, {Sivarani}, {Reddy}, \& {Rajaguru}}]{Swastik2021}
{Swastik}, C., {Banyal}, R.~K., {Narang}, M., {et~al.} 2021, \aj, 161, 114

\bibitem[{{Tennyson} {et~al.}(2020){Tennyson}, {Yurchenko}, {Al-Refaie}, {Clark}, {Chubb}, {Conway}, {Dewan}, {Gorman}, {Hill}, {Lynas-Gray}, {Mellor}, {McKemmish}, {Owens}, {Polyansky}, {Semenov}, {Somogyi}, {Tinetti}, {Upadhyay}, {Waldmann}, {Wang}, {Wright}, \& {Yurchenko}}]{Tennyson2020}
{Tennyson}, J., {Yurchenko}, S.~N., {Al-Refaie}, A.~F., {et~al.} 2020, \jqsrt, 255, 107228

\bibitem[{{Tremblin} {et~al.}(2016){Tremblin}, {Amundsen}, {Chabrier}, {Baraffe}, {Drummond}, {Hinkley}, {Mourier}, \& {Venot}}]{Tremblin2016}
{Tremblin}, P., {Amundsen}, D.~S., {Chabrier}, G., {et~al.} 2016, \apjl, 817, L19

\bibitem[{{Tremblin} {et~al.}(2015){Tremblin}, {Amundsen}, {Mourier}, {Baraffe}, {Chabrier}, {Drummond}, {Homeier}, \& {Venot}}]{Tremblin2015}
{Tremblin}, P., {Amundsen}, D.~S., {Mourier}, P., {et~al.} 2015, \apjl, 804, L17

\bibitem[{{Turrini} {et~al.}(2021){Turrini}, {Schisano}, {Fonte}, {Molinari}, {Politi}, {Fedele}, {Pani{\'c}}, {Kama}, {Changeat}, \& {Tinetti}}]{Turrini2021}
{Turrini}, D., {Schisano}, E., {Fonte}, S., {et~al.} 2021, \apj, 909, 40

\bibitem[{Virtanen {et~al.}(2020)Virtanen, Gommers, Oliphant, Haberland, Reddy, Cournapeau, Burovski, Peterson, Weckesser, Bright, {van der Walt}, Brett, Wilson, Millman, Mayorov, Nelson, Jones, Kern, Larson, Carey, Polat, Feng, Moore, {VanderPlas}, Laxalde, Perktold, Cimrman, Henriksen, Quintero, Harris, Archibald, Ribeiro, Pedregosa, {van Mulbregt}, \& {SciPy 1.0 Contributors}}]{Scipy2020}
Virtanen, P., Gommers, R., Oliphant, T.~E., {et~al.} 2020, Nature Methods, 17, 261

\bibitem[{{Visser} {et~al.}(2009){Visser}, {van Dishoeck}, \& {Black}}]{Visser2009}
{Visser}, R., {van Dishoeck}, E.~F., \& {Black}, J.~H. 2009, \aap, 503, 323

\bibitem[{{Wakeford} {et~al.}(2017){Wakeford}, {Visscher}, {Lewis}, {Kataria}, {Marley}, {Fortney}, \& {Mandell}}]{Wakeford2017}
{Wakeford}, H.~R., {Visscher}, C., {Lewis}, N.~K., {et~al.} 2017, \mnras, 464, 4247

\bibitem[{{Wang} {et~al.}(2020){Wang}, {Tennyson}, \& {Yurchenko}}]{Wang2020}
{Wang}, Y., {Tennyson}, J., \& {Yurchenko}, S.~N. 2020, Atoms, 8, 7

\bibitem[{{Wilson}(1999)}]{Wilson1999}
{Wilson}, T.~L. 1999, Reports on Progress in Physics, 62, 143

\bibitem[{{Winn} {et~al.}(2017){Winn}, {Petigura}, {Morton}, {Weiss}, {Dai}, {Schlaufman}, {Howard}, {Isaacson}, {Marcy}, {Justesen}, \& {Albrecht}}]{Winn2017}
{Winn}, J.~N., {Petigura}, E.~A., {Morton}, T.~D., {et~al.} 2017, \aj, 154, 270

\bibitem[{{Wu} {et~al.}(2020){Wu}, {Bowler}, {Sheehan}, {Andrews}, {Herczeg}, {Kraus}, {Ricci}, {Wilner}, \& {Zhu}}]{Wu2020}
{Wu}, Y.-L., {Bowler}, B.~P., {Sheehan}, P.~D., {et~al.} 2020, \aj, 159, 229

\bibitem[{{Wu} {et~al.}(2017){Wu}, {Close}, {Eisner}, \& {Sheehan}}]{Wu2017}
{Wu}, Y.-L., {Close}, L.~M., {Eisner}, J.~A., \& {Sheehan}, P.~D. 2017, \aj, 154, 234

\bibitem[{{Xuan} {et~al.}(2024{\natexlab{a}}){Xuan}, {Hsu}, {Finnerty}, {Wang}, {Ruffio}, {Zhang}, {Knutson}, {Mawet}, {Mamajek}, {Inglis}, {Wallack}, {Bryan}, {Blake}, {Molli{\`e}re}, {Hejazi}, {Baker}, {Bartos}, {Calvin}, {Cetre}, {Delorme}, {Doppmann}, {Echeverri}, {Fitzgerald}, {Jovanovic}, {Liberman}, {L{\'o}pez}, {Morris}, {Pezzato}, {Sappey}, {Schofield}, {Skemer}, {Wallace}, {Wang}, {Agrawal}, \& {Horstman}}]{Xuan2024}
{Xuan}, J.~W., {Hsu}, C.-C., {Finnerty}, L., {et~al.} 2024{\natexlab{a}}, \apj, 970, 71

\bibitem[{{Xuan} {et~al.}(2024{\natexlab{b}}){Xuan}, {Wang}, {Finnerty}, {Horstman}, {Grimm}, {Peck}, {Nielsen}, {Knutson}, {Mawet}, {Isaacson}, {Howard}, {Liu}, {Walker}, {Phillips}, {Blake}, {Ruffio}, {Zhang}, {Inglis}, {Wallack}, {Sanghi}, {Gonzales}, {Dai}, {Baker}, {Bartos}, {Bond}, {Bryan}, {Calvin}, {Cetre}, {Delorme}, {Doppmann}, {Echeverri}, {Fitzgerald}, {Jovanovic}, {Liberman}, {L{\'o}pez}, {Martin}, {Morris}, {Pezzato}, {Ruane}, {Sappey}, {Schofield}, {Skemer}, {Venenciano}, {Wallace}, {Wang}, {Wizinowich}, {Xin}, {Agrawal}, {Do {\'O}}, {Hsu}, \& {Phillips}}]{Xuan2024b}
{Xuan}, J.~W., {Wang}, J., {Finnerty}, L., {et~al.} 2024{\natexlab{b}}, \apj, 962, 10

\bibitem[{{Yoshida} {et~al.}(2022){Yoshida}, {Nomura}, {Furuya}, {Tsukagoshi}, \& {Lee}}]{Yoshida2022}
{Yoshida}, T.~C., {Nomura}, H., {Furuya}, K., {Tsukagoshi}, T., \& {Lee}, S. 2022, \apj, 932, 126

\bibitem[{{Zhang} {et~al.}(2024){Zhang}, {Gonz{\'a}lez Picos}, {de Regt}, {Snellen}, {Gandhi}, {Ginski}, {Kesseli}, {Landman}, {Molli{\`e}re}, {Nasedkin}, {S{\'a}nchez-L{\'o}pez}, {Stolker}, {Inglis}, {Knutson}, {Mawet}, {Wallack}, \& {Xuan}}]{Zhang2024}
{Zhang}, Y., {Gonz{\'a}lez Picos}, D., {de Regt}, S., {et~al.} 2024, \aj, 168, 246

\bibitem[{{Zhang} {et~al.}(2021{\natexlab{a}}){Zhang}, {Snellen}, {Bohn}, {Molli{\`e}re}, {Ginski}, {Hoeijmakers}, {Kenworthy}, {Mamajek}, {Meshkat}, {Reggiani}, \& {Snik}}]{Zhang2021}
{Zhang}, Y., {Snellen}, I. A.~G., {Bohn}, A.~J., {et~al.} 2021{\natexlab{a}}, \nat, 595, 370

\bibitem[{{Zhang} {et~al.}(2021{\natexlab{b}}){Zhang}, {Snellen}, \& {Molli{\`e}re}}]{Zhang2021_BD}
{Zhang}, Y., {Snellen}, I. A.~G., \& {Molli{\`e}re}, P. 2021{\natexlab{b}}, \aap, 656, A76

\bibitem[{{Zhang} {et~al.}(2023){Zhang}, {Molli{\`e}re}, {Hawkins}, {Manea}, {Fortney}, {Morley}, {Skemer}, {Marley}, {Bowler}, {Carter}, {Franson}, {Maas}, \& {Sneden}}]{Zhang2023}
{Zhang}, Z., {Molli{\`e}re}, P., {Hawkins}, K., {et~al.} 2023, \aj, 166, 198

\bibitem[{{Zhou} {et~al.}(2014){Zhou}, {Herczeg}, {Kraus}, {Metchev}, \& {Cruz}}]{Zhou2014}
{Zhou}, Y., {Herczeg}, G.~J., {Kraus}, A.~L., {Metchev}, S., \& {Cruz}, K.~L. 2014, \apjl, 783, L17

\end{thebibliography}

\begin{appendix}\onecolumn

\section{Full best-fit models} \label{app:bestfit}
\begin{figure*}[ht!]
    \centering
    \includegraphics[width=0.9\linewidth]{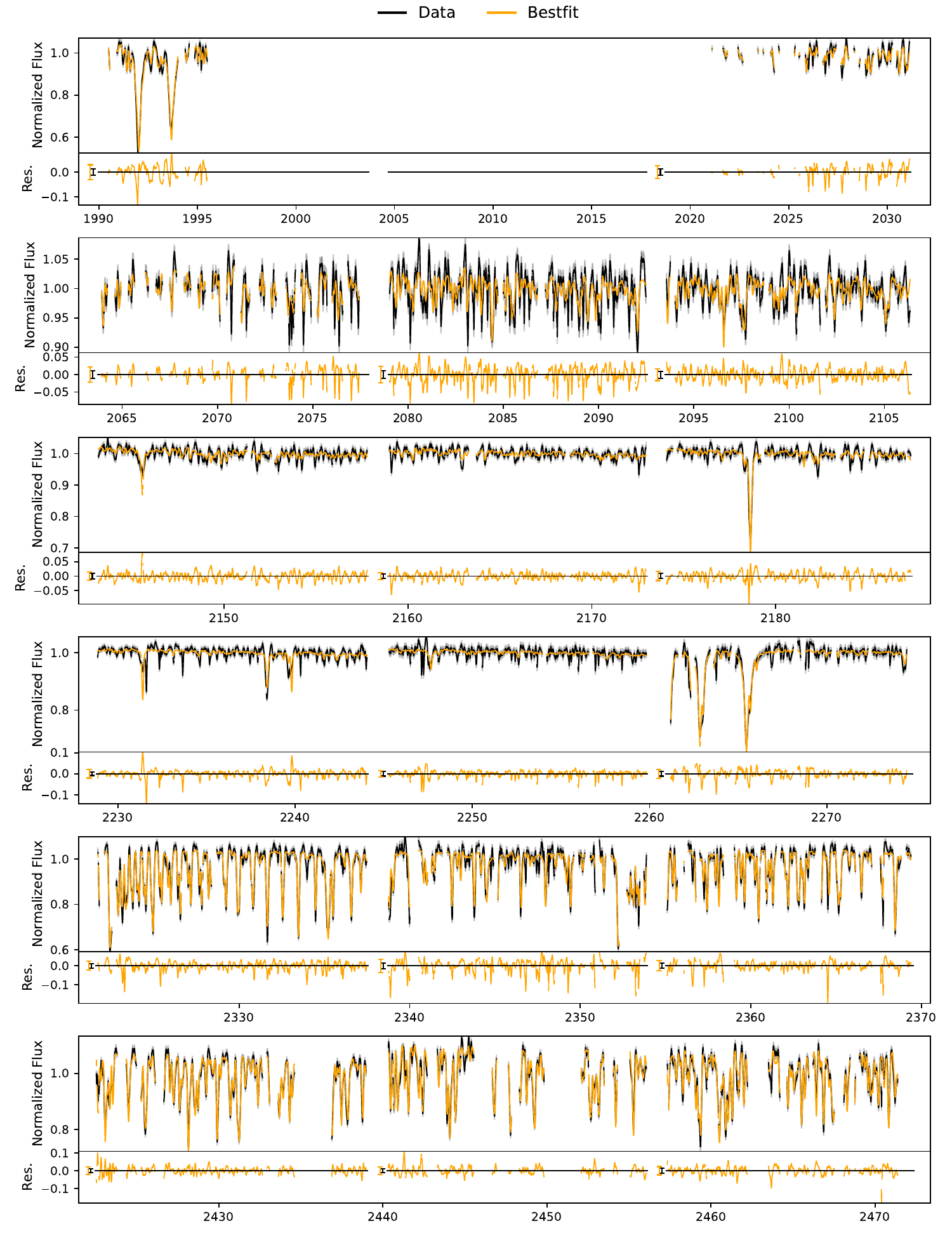} 
    \caption{Best-fit model for ROXs~12A, with each panel showing a different spectral order. The observed spectrum is shown in black, along with its 1\,$\sigma$ uncertainties in gray, and the best-fit model in yellow. In the panel underneath each order, we show the residuals (data minus model), as well as the data uncertainties for each order-detector pair as a black error bar and the standard deviation of the residuals as a yellow error bar.}\label{fig:bestfit_split_A}
\end{figure*}

\begin{figure*}[ht!]
    \centering
    \includegraphics[width=0.9\linewidth]{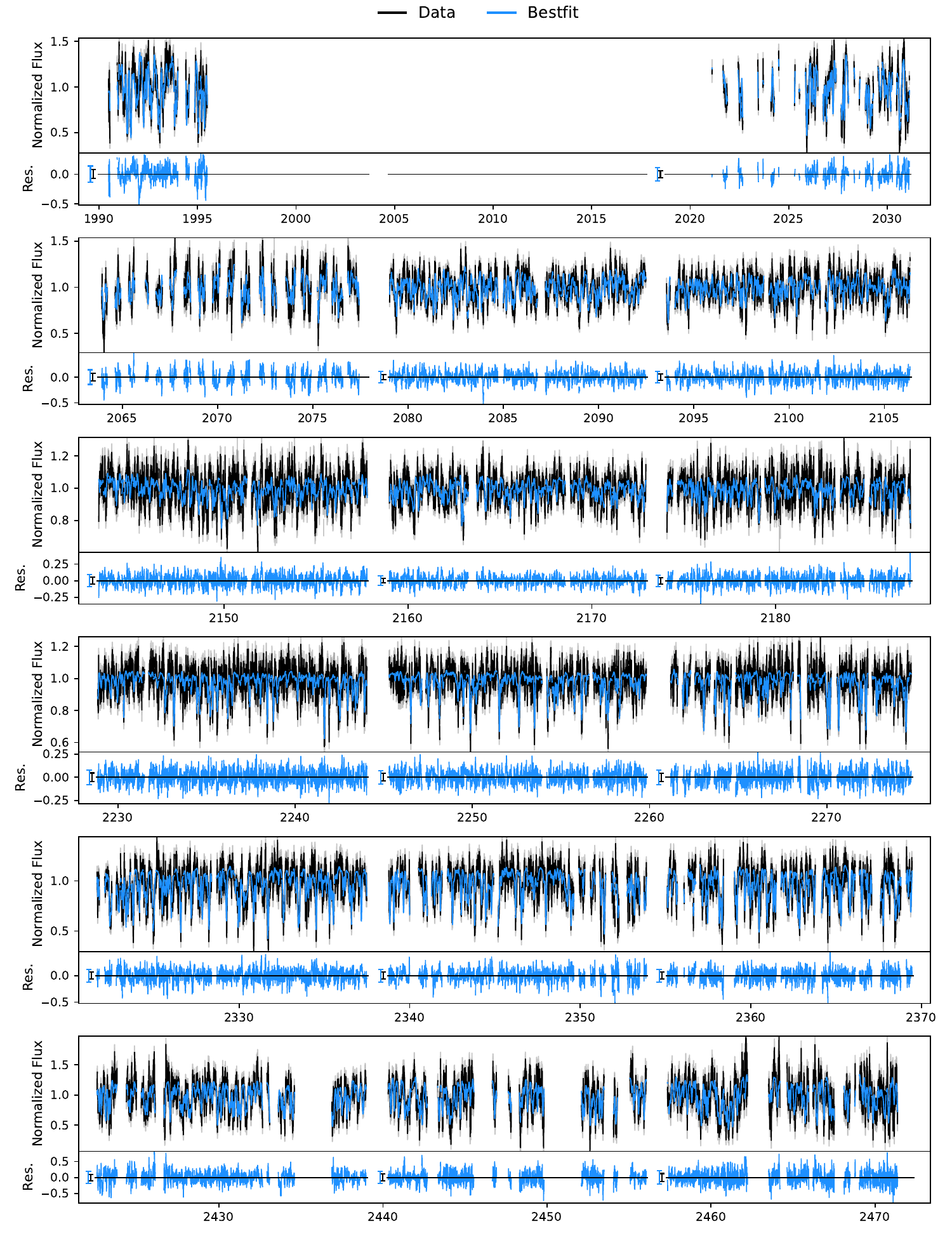}
    \caption{Best-fit model for ROXs~12B, with each panel showing a different spectral order. The observed spectrum is shown in blue, along with its 1\,$\sigma$ uncertainties in gray, and the best-fit model in blue. In the panel underneath each order, we show the residuals (data minus model), as well as the data uncertainties for each order-detector pair as a black error bar and the standard deviation of the residuals as a blue error bar.}
\end{figure*}

\clearpage
\section{Injected test spectrum} \label{app:inj}

\begin{figure}[ht!]
    \centering
    \includegraphics[width=0.45\textwidth]{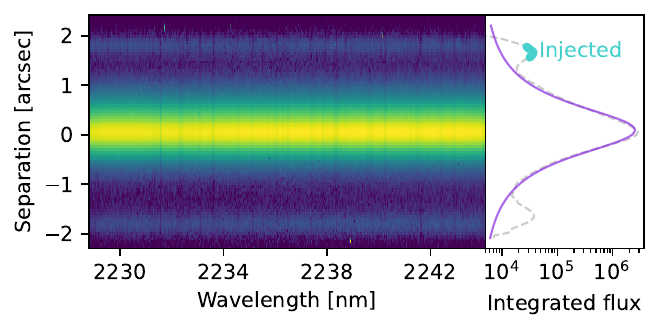}
    \caption{Equivalent to Fig.~\ref{fig:observation}, except for the addition of an injected synthetic spectrum on the other side of the stellar PSF at -1.7\,arcsec. The spectral extraction follows the same steps as described for the true companion (see Sect.~\ref{sec:observations}).}
    \label{fig:obs_inj}
\end{figure}

\begin{figure*}[ht!]
    \centering
    \includegraphics[width=0.855\linewidth]{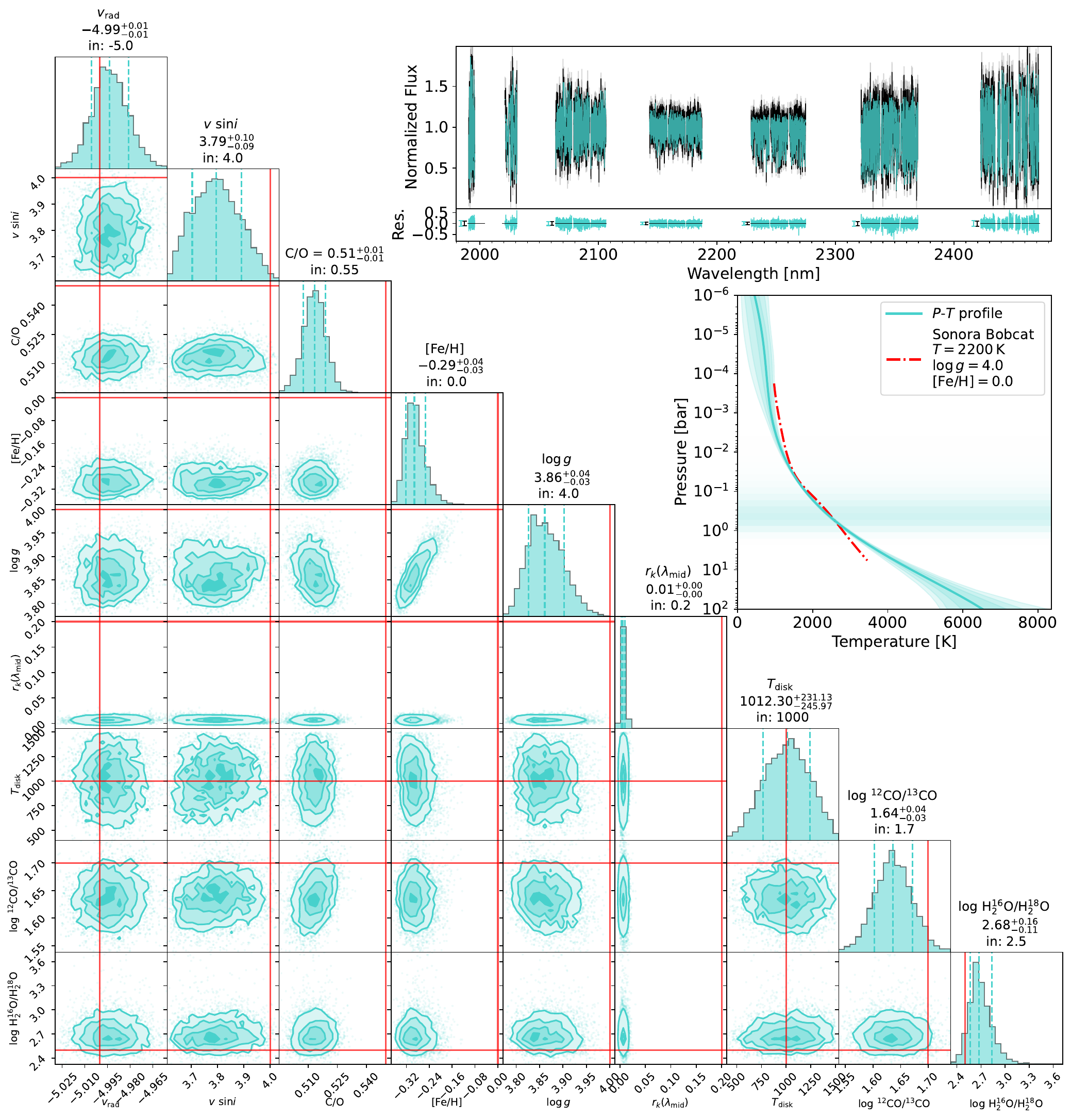}
    \caption{Summary of the retrieval results of the injected test spectrum. We show the Sonora Bobcat model $P$--$T$ profile that was used as an input for generating the spectrum. The red lines in the cornerplot represent the input values.}
    \label{fig:summary_inj}
\end{figure*}

\clearpage
\section{ROXs 12A cross-correlation functions} \label{app:ccf_A}

\begin{figure}[ht!]
    \centering\includegraphics[width=0.45\textwidth]{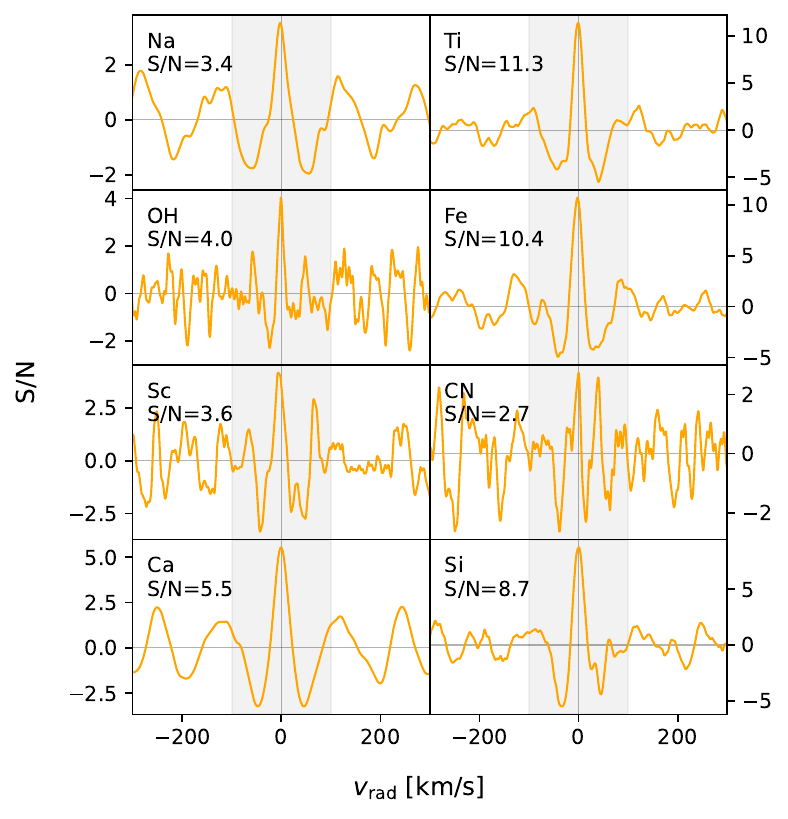}
    \caption{Cross-correlation functions of the remaining atmospheric species in ROXs~12A.}
    \label{fig:ccf_rest_A}
\end{figure}

\end{appendix}

\end{document}